\documentclass{aastex631}

\renewcommand{\thefootnote}{\arabic{footnote}}
\usepackage[perpage]{footmisc}

\begin{document}

\title{The EMPI Code for Plasma-Induced Effects on Radio Waves \uppercase\expandafter{\romannumeral1}: Non-Magnetized Media and Applications to Fast Radio Bursts
}

\author{Nan Xu}
\altaffiliation{nan.xu@mail.bnu.edu.cn}
\affiliation{School of Physics and Astronomy, Beijing Normal University, Beijing 100875, China}
\affiliation{Institute for Frontier in Astronomy and Astrophysics, Beijing Normal University, Beijing 102206, China}

\author{He Gao}
\altaffiliation{gaohe@bnu.edu.cn}
\affiliation{School of Physics and Astronomy, Beijing Normal University, Beijing 100875, China}
\affiliation{Institute for Frontier in Astronomy and Astrophysics, Beijing Normal University, Beijing 102206, China}


\author{Yuan-Pei Yang}
\altaffiliation{ypyang@ynu.edu.cn}
\affiliation{South-Western Institute For Astronomy Research, Yunnan University, Yunnan 650504, China}
\affiliation{Purple Mountain Observatory, Chinese Academy of Sciences, Nanjing 210023, China}

\author{Bing Zhang}
\affiliation{Nevada Center for Astrophysics, University of Nevada, Las Vegas, NV 89154, USA}
\affiliation{Department of Physics and Astronomy, University of Nevada, Las Vegas, NV 89154, USA}

\author{Wei-Yang Wang}
\affiliation{School of Astronomy and Space Science, University of Chinese Academy of Sciences, Beijing 100049, China}

\author{Tian-Cong Wang}
\affiliation{School of Physics and Astronomy, Beijing Normal University, Beijing 100875, China}
\affiliation{Institute for Frontier in Astronomy and Astrophysics, Beijing Normal University, Beijing 102206, China}

\author{Ran Gao}
\affiliation{School of Physics and Astronomy, Beijing Normal University, Beijing 100875, China}
\affiliation{Institute for Frontier in Astronomy and Astrophysics, Beijing Normal University, Beijing 102206, China}

\begin{abstract}
Electromagnetic waves undergo modifications as they propagate through plasma. 
We present electromagnetic wave plasma interaction (EMPI), a three-dimensional numerical framework designed to simulate the interaction between radio signals and cold plasma. 
With input plasma density profiles, intrinsic radio signals, and the time and frequency resolutions of the telescope, the code synthesizes observed signals using first-principles calculations. 
EMPI is capable of modeling a wide range of plasma distributions, spanning analytically described smooth functions (e.g., Gaussian or exponential profiles), statistical models (e.g., turbulent screens), and discrete macroscopic structures like isolated plasma clumps, which are difficult to model both analytically and statistically. 
Validation tests demonstrate excellent agreement with established plasma propagation effects, such as dispersion, lensing, scintillation, and scattering. 
This code provides an efficient method for handling both analytical and statistical scenarios, bridging the gap between these descriptions. 
Thanks to its comprehensive capabilities, EMPI is particularly useful for studying radio sources of cosmological origin, especially pulse-like signals such as fast radio bursts. 
As these signals travel through diverse and complex plasma environments across the universe, their properties are inevitably altered, resulting in observable changes. 
In this context, EMPI serves as a valuable tool for studying the propagation effects of these sources, helping to advance the understanding of their essence and the intervening plasma environments.

\end{abstract}


\section{Introduction} \label{sec:intro}
 Fast radio bursts (FRBs) emerged as a key focus in contemporary astronomy since the discovery of the first FRB exhibiting excess dispersion measure (DM) \citep{Lorimer}. 
Subsequent studies \citep{Thornton2013, Tendulkar_2017}  have confirmed their extragalactic origins, establishing the sources at cosmological distances.
Propagating over such enormous distances, these signals inevitably traverse diverse plasma media that modify their observable characteristics. 
Understanding these propagation effects that cause deviations from the intrinsic characteristics of the bursts is crucial both for advancing the knowledge of FRB mechanisms, which remain under debate  \citep{zhang_physics_2023}, and for probing the nature of intervening plasma \citep{Zheng_2014}. 

Numerous studies have systematically explored the impact of plasma on FRB signals, successfully accounting for key propagation phenomena including dispersion  \citep{McQuinn_2014, Yang_2017, kulkarni2020, Macquart2020}, Faraday rotation and conversion \citep{FengYi_Science, Qu_Propagation, YangXuZhang2023, XiaYnagLi_2023,wang2025}, plasma lensing \citep{Clegg_1998, Cordes_2017, Er_2020}, scintillation \citep{Rickett1977, Rickett1990, KumarConstrainMechnism} and scattering \citep{Luan_2014, cordes2016, Xu_2016}. 
Building on these foundations, recent observational and theoretical advances have further expanded our understanding of these effects.
Observations unveiled novel signatures: wideband propagation effects \citep{PropagationEffectsLowFrequencies}, precise timescale measurements \citep{TimeScale20201124A}, scattering variability \citep{ScatteringVariability}, and scintillation arc morphologies \citep{scintillationArc201124A,scintillationArc220912A}. 
Concurrently, refined propagation models now quantitatively explain FRB-related phenomena, from burst enhancement or suppression \citep{ChenXuechun}, scattering-induced pulse broadening \citep{BroadeningPulseWidthFRB}, to spectral changes \citep{Levkov_2022_spectral, Kumar_2024_spectral}.
Such theoretical progress has transformed propagation effects into diagnostic tools for both sight line plasma characterization \citep{Prochaska2019, shin2024InvestigatingSightLine} and radiation mechanism constraints \citep{KumarConstrainMechnism, nimmo2024KumarObservation}.
This evolution from qualitative estimation to quantitative modeling \citep{TwoScreen,ScintillationModelling} enables more precise interpretations.
However, it exposed critical limitations: Current approaches predominantly rely on idealized plasma configurations, i.e., whether analytic density profiles (e.g., Gaussian/exponential lenses) or statistical turbulence (e.g., Kolmogorov-type screens). 
Such simplifications starkly contrast with realistic plasma structures exhibiting unmodeled complexity in both spatial configuration and temporal evolution.
Where analytical descriptions fail and statistical assumptions break down (configurations lacking analytical descriptions and sufficient randomness for statistical treatment), elaborate numerical modeling becomes imperative for accurate quantification.

To bridge this gap, we present EMPI, a three-dimensional (3D) numerical framework based on first principles, enabling calculations of electromagnetic wave propagation through arbitrarily structured plasma media. Validated against established theories, EMPI holistically captures comprehensive propagation effects rather than isolating individual phenomena, enabling precise modeling of noncanonical plasma geometries. This versatility facilitates critical constraints on plasma properties imprinting on FRB signals during cosmic traversal.

Crucially, the urgency for such advanced modeling tools is amplified by impending observational revolutions. The Five
hundred-meter Aperture Spherical Radio Telescope, Square Kilometre Array, and other large-scale facilities are poised to detect fainter bursts and subtle plasma imprints and map dynamic structural evolution. 
These observational capacities would remain underexploited without corresponding advances in plasma interaction modeling. 
EMPI serves as an essential tool for investigating the propagation effects of FRBs, supporting future studies on novel observations.
Beyond FRBs, the framework extends naturally to plasma-induced effects studies of other radio emitters such as pulsars and persistent radio sources.

As the inaugural work in the EMPI code series, this paper establishes the foundational framework for nonmagnetized plasma regimes, comprehensively addressing all key propagation effects: dispersion, plasma lensing, scintillation, and scattering. 
Deliberately excluding magnetized plasma and relativistic, nonlinear effects (see dedicated discussions in \citet{Luan_2014, KumarLu,Yang_2020}), this work focuses on developing first-principles solutions for linear propagation effects. Subsequent studies will seamlessly integrate magnetic field effects and explore advanced scientific applications.

This paper is structured as follows: Section~\ref{sec:Theoretical_Background} provides the theoretical background of plasma interactions with FRBs;
Section~\ref{sec:Code} details the numerical architecture with modular implementation strategies; 
Section~\ref{sec:Results} validates the code through (i) benchmark tests against analytical solutions and (ii) demonstrations of complex scenario handling; 
Section~\ref{sec:Summary} summarizes the performance of the code while charting future research directions, focusing on (i) refining plasma lensing and scattering models and (ii) developing parameter constraint methods for observational studies.

\section{Code Principles and Methodology} \label{sec:Theoretical_Background}
The plasma–FRB interaction is inherently multifaceted.
For clarity, this study restricts its scope to nonmagnetized plasma regimes, where time delays, phase shifts, and the electron encountered collectively manifest as four dominant effects: dispersion, plasma lensing, scintillation, and scattering. 
This focus is justified, since nonmagnetized plasma interactions not only pervade astrophysical environments but also constitute the necessary foundation for probing intricate magnetized plasma effects.

\begin{figure}[ht!] 
\centering 
\includegraphics[width=0.8\textwidth]{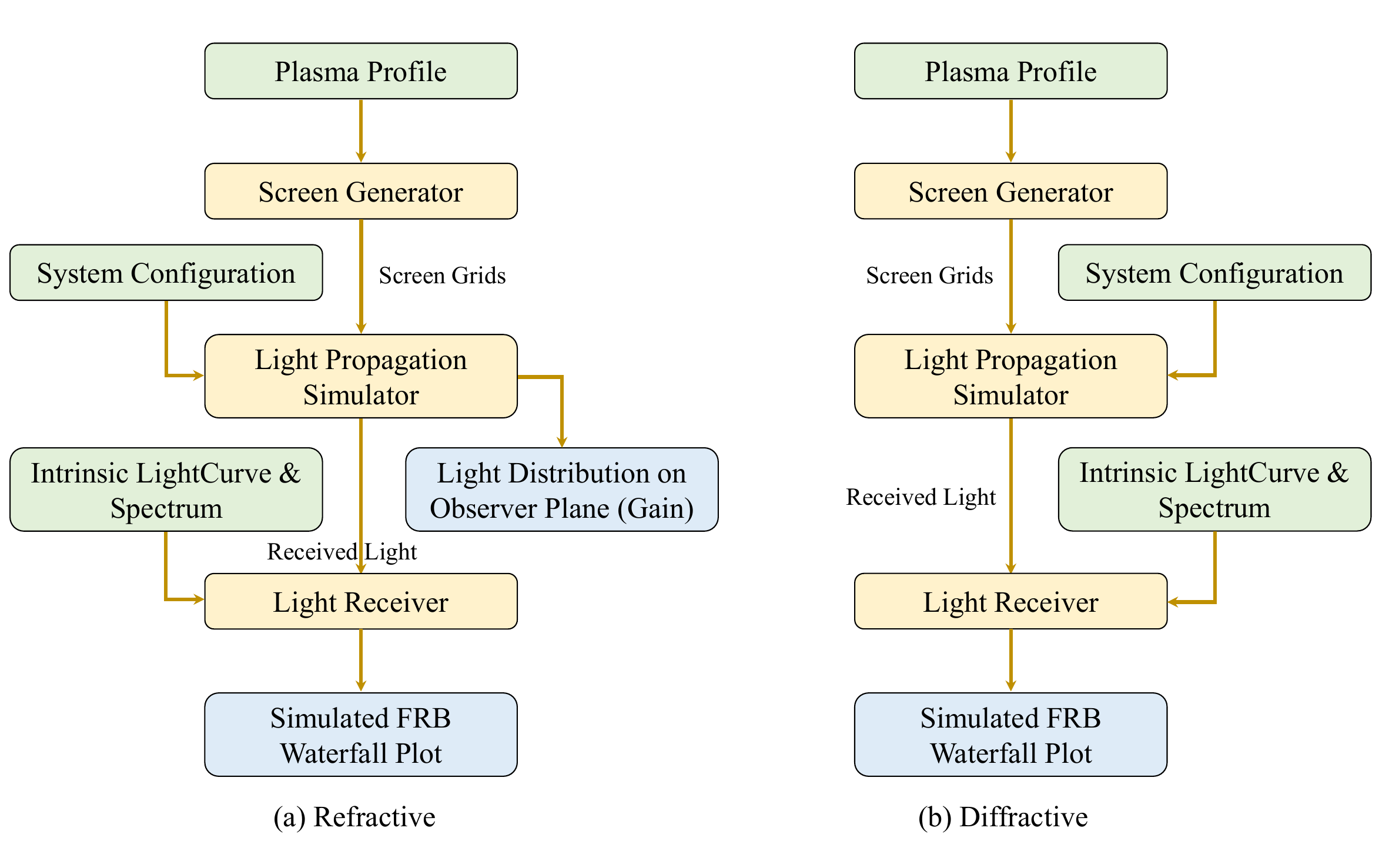} 
\caption{Schematic workflow of the EMPI framework for simulating FRB signal propagation through plasma media. (a) Refractive Regime: Macro-scale plasma density gradients are resolved using geometric optics. (b) Diffractive Regime:  Small-scale turbulent structures are modeled via a statistical description. Color coding: Green rectangles indicate inputs, yellow regions represent multi-stage computational modules, and blue areas denote outputs.
}
\label{Fig:FlowChart} 
\end{figure}

This section establishes the theoretical underpinnings of the EMPI framework for modeling plasma-induced propagation effects. 
The computational workflow operates in two regimes, refractive and diffractive, determined by characteristic plasma spatial scales. 
First, we introduce the criteria for regime selection, followed by a detailed description of the core processing steps in each regime. 
Next, additional treatments applied to both regimes are outlined. 
Finally, regardless of the regime, we present the method for processing light to construct the received FRB signal.
The overall code workflow is illustrated in Figure~\ref{Fig:FlowChart}, with a detailed explanation of each module provided in Section~\ref{sec:Code}.

\subsection{Criteria for Regime Selection}\label{Sec_Regime_selection}
Plasma-induced plasma-EM wave interactions alter the propagation direction of light via two distinct mechanisms: refraction and diffraction. 
Refraction occurs when plasma structures are significantly larger than the signal wavelength, permitting a geometric optics approximation.
In contrast, diffraction dominates for sub-wavelength plasma inhomogeneities, necessitating a wave-optics treatment based on the Fresnel–Kirchhoff integral. 
A detailed discussion of these regimes is provided in \citet{Jow_Pen_2023}.
Fundamentally, the choice between refractive and diffractive treatment hinges on the relative size of the plasma structures compared to the signal wavelength. 
In this paper, we derive the criterion by comparing the refractive and diffractive deflection angles caused by the same plasma clump, thereby determining which effect is dominant \citep{draine_physics_2011} (see Appendix \ref{criteria} for further details). 
The critical size $l_{\rm c}$ is defined as:
\begin{equation}
    l_{\rm c} = \frac{2\pi m_{\rm e}\nu c}{q^{2}\Delta n},
\end{equation}
where $m_{\rm e}$ is the electron mass, $\nu$ is the frequency, $c$ is the speed of light in the vacuum, $q$ is the electron charge, and $\Delta n$ represents the characteristic electron density fluctuation measured at the scale of the plasma clump.
This critical size delineates the transition between refractive and diffractive regimes: when the physical plasma bump size $l$ is much greater than $l_{\rm c}$, refraction dominates, conversely, when $l \ll l_{\rm c}$, diffraction prevails.  
This criterion establishes theoretical foundations for distinguishing dominant physical mechanism and informs the selection of appropriate computational methodologies tailored to distinct regimes.

Physically, the refractive regime corresponds to plasma lensing and other relatively large-scale or continuously varying structures, while the diffractive regime represents turbulent scattering screens with small-scale variations. 
Although this distinction is clear in principle, intermediate cases, such as configurations with several large plasma clumps, may not fit neatly into either category. 
In these instances, determining the dominant effect relies on the criteria outlined above, and users must carefully assess the input plasma profile to select the appropriate computational regime.

The characteristic physical bump size is an intrinsic property of the input plasma profile, theoretically spanning arbitrary magnitudes. 
However, numerical calculations introduce resolution-dependent constraints. The plasma screen is described by discrete grid points in 3D space, with the density precisely defined at each point. 
Grid spacing establishes the minimum resolvable feature size. 
A bump must occupy multiple adjacent nodes to form discernible structures.
Consequently, the physical size of a plasma bump cannot be smaller than the grid resolution.

\subsection{Refractive Regime}\label{Refractive Section}
In conventional geometric optics, the ray-based approach to studying light propagation and redistribution inherently neglects phase information, as it is not crucial for incoherent light. 
Nevertheless, when applied to radio waves, especially coherent signals, phase information becomes significant. 
In this method, phase shifts are calculated alongside ray trajectories, allowing geometric optics to not only trace ray paths but also capture phase variations along each trajectory. 
This integration ensures accurate phase accounting, which is essential for analyzing coherent signals.

Within this framework, the refractive regime models light as rays originating from a point source, where each ray propagates independently along a trajectory perpendicular to the local wave front surface. 
When light propagates through a plasma screen, spatial variations in the plasma density generate position-dependent phase shifts. 
These shifts distort the wave front surface and induce deflections in the light propagation direction. 
To computationally quantify this effect, the plasma screen is divided into 3D discrete grid points. 
As the light penetrates the screen, phase shifts are determined by the plasma density.

The phase shift $\phi$ consists of two primary contributions: the geometric component (first term in Eq.~\ref{PhaseShift}), originating from the distance traveled along the ray trajectory, and the dispersive component (second term in Eq.~\ref{PhaseShift}), resulting from the interaction with the non-magnetized plasma.
This relationship is described by the following equation:
\begin{equation}\label{PhaseShift}
    \phi  = \int \frac{\omega}{c} \left( 1 - \frac{\omega^{2}_{\rm p}}{2\omega^{2}} \right) ds,
\end{equation}
which is derived from the plasma dispersion relation (see Appendix \ref{criteria} for details), where $\omega$ is the angular frequency and $\omega_{\rm p}$
is the plasma frequency. 
The plasma frequency $\omega_{\rm p}$ can be expressed as $\omega_{\rm p} = \sqrt{4\pi q^2n_{\rm e}/m_{\rm e}}$, where $n_{\rm e}$ is the number density of electrons. 
With the phase shift known, the new local wave front can be obtained, and principal component analysis (PCA)\footnote{PCA is a statistical method that identifies the directions (or principal components) along which data varies the most. 
The process involves computing the eigenvectors and eigenvalues of the covariance matrix of the data points. 
These eigenvectors represent the axes of maximum variance, and the eigenvalue magnitudes indicate the strength of variance along each axis. 
The axis with the smallest eigenvalue, indicating the least variation, is taken as the normal direction of the wave front.} 
is used to find the normal vectors, indicating the new propagation direction of the light.

During the passage through the plasma screen, their behavior is computed via a layer-wise iterative process along the main propagation direction (the x-direction in Cartesian coordinates, as described in Section~\ref{sec:Code}). 
This approach enables a detailed analysis of interactions between light and the inhomogeneous plasma media.

\begin{figure}[ht!] 
\centering 
\includegraphics[width=0.8\textwidth]{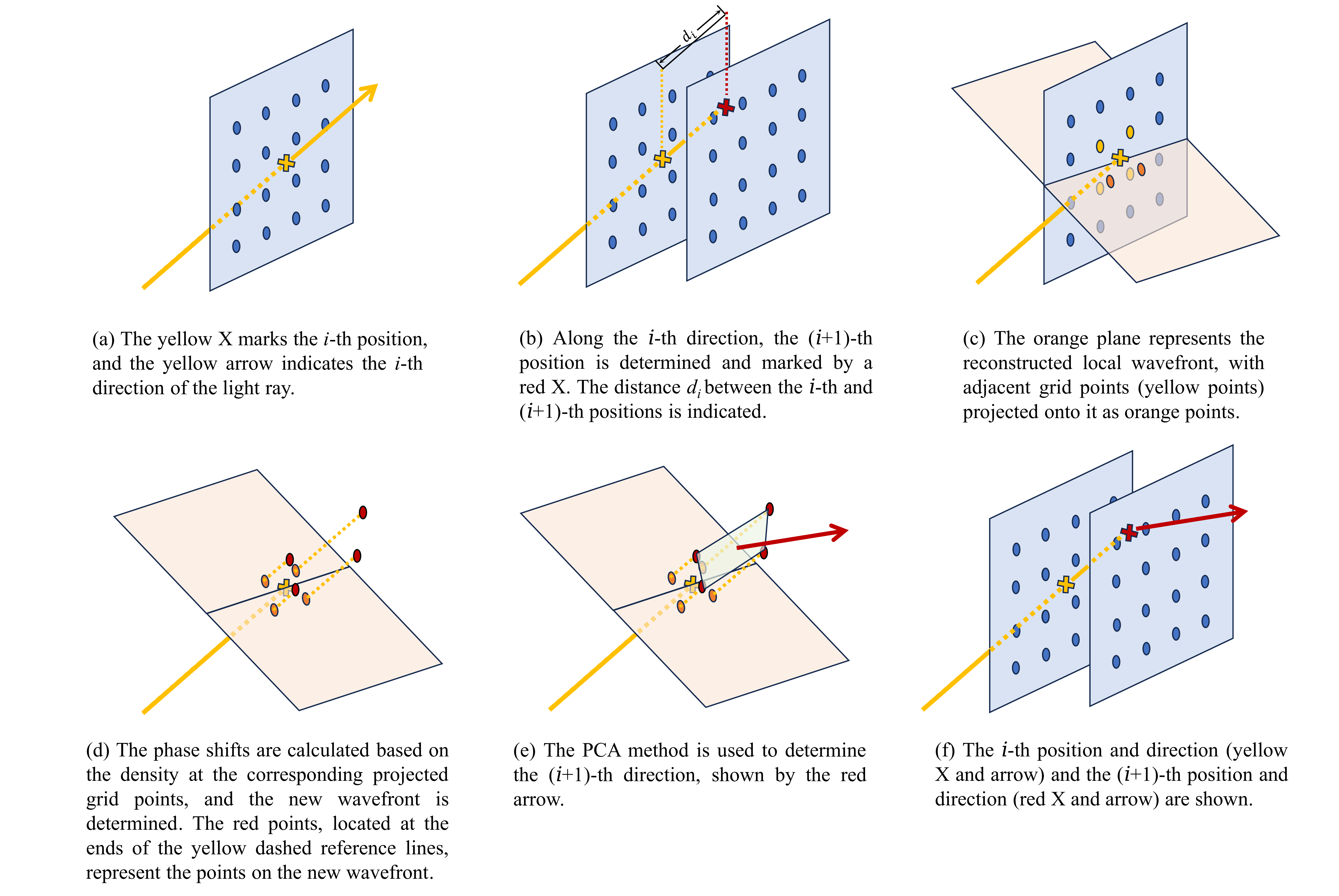} 
\caption{Schematic representation of the iterative process for light propagation through a layered plasma screen.}
\label{Fig:Iteration} 
\end{figure}

The iteration process is delineated in detail in Figure~\ref{Fig:Iteration}.
It begins at the incident point on the plasma screen, where the initial direction of each ray is determined by the vector connecting the source to its intersection point on the screen.
Due to the astronomical distance between the source and the screen, these vectors are nearly parallel across different rays, with only small deviations. 
The incident position and initial direction set the first step of the iteration and provide the starting conditions for subsequent calculations. 
In the $i$-th step (layer), the incoming direction and position of the light ray are used to calculate the intersection point with the next layer, defining the position of the ($i$+1)-th step (see panel (b) in Figure~\ref{Fig:Iteration}). 
The distance between these two points $d_i$ represents the distance traveled by the light in the $i$-th layer. 
This distance is subsequently used to calculate travel time $t_i$, phase shift $\phi_i$, and the number of electrons encountered or dispersion measure ${\rm DM}_i$, for this layer.

To compute the updated local wave front at each intersection point, the position and density information from its $n$ surrounding grid points, where the density is precisely defined, are utilized. 
A minimum of three points is required for the calculation; however, four points are used in this implementation to avoid numerical instabilities. 
The incident direction allows for the reconstruction of an initial local wave front plane (orange plane in panel (c) of Figure~\ref{Fig:Iteration}).
The four surrounding points are projected onto this wave front plane (orange points in panel (c) of Figure~\ref{Fig:Iteration}).
The key is to identify the four corresponding new wavefront points. 
These four surrounding points yield four distinct phase shift values, calculated from identical distances $d_{i}$ but different densities, via Eq.~\ref{PhaseShift}. 
Then the four phase shift values are converted into spatial
displacements using $\Delta_{\phi} = \phi\lambda /2\pi$, where $\lambda$ is the wavelength. 
The positions of the updated wave front points in 3D space are obtained by offsetting the initial wave front points on the original wave front plane by their respective spatial distances $\Delta_{\phi}$ along the $i$-th direction. (see panel (d) in Figure~\ref{Fig:Iteration}).
Finally, the normal vector direction of the updated local wave front is determined by applying the PCA method to these four updated wave front points, defining the propagation direction for the ($i$+1)-th step (see panel (e) in Figure~\ref{Fig:Iteration}).

In each step, relevant physical parameters, such as travel time $t_{i}$, phase shift $\phi_{i}$ and dispersion measure ${\rm DM}_{i}$, are computed as well. 
The density at the intersection point in the $i$-th step $n_{{\rm e}, i}$ is obtained by interpolating the density of the surrounding grid points. 
Specifically, the travel time $t_{i}$ is given by the following relation: 
\begin{equation}
    t_{i} = \frac{d_{i}}{v_{{\rm g,}i}} \approx (1+\frac{1}{2}\frac{\omega_{{\rm p,}i}^{2}}{\omega^{2}})\frac{d_{i}}{c},
\end{equation}
where $v_{{\rm g,}i}$ is the group velocity of the wave in the plasma and the second approximation comes from the condition $\omega \gg \omega_{\rm p}$ \citep{zhang_physics_2023}. 
The phase shift $\phi_{i}$ follows Eq.~\ref{PhaseShift},
\begin{equation}
     \phi_{i}  =  \frac{\omega d_{i}}{c} \left( 1 - \frac{\omega^{2}_{{\rm p,}i}}{2\omega^{2}} \right).
\end{equation}
And the dispersion measure ${\rm DM}_{i}$ is computed as:
\begin{equation}
    {\rm DM}_i = n_{{\rm e,}i}d_{i}.
\end{equation}
Upon completion of one iteration cycle, the accumulated time, phase, and DM values are stored for subsequent computational stages.

The iterative procedure numerically determines the final propagation vectors and spatial coordinates of the light rays at the plasma screen exit plane. 
To evaluate observational detectability, the trajectory of each ray is analyzed using the derived exit parameters and the predefined separation distance $d_{\rm po}$ between the plasma screen and the observer plane.
These parameters enable direct determination of the intersection point of each ray with the observer plane. 
By comparing these intersection coordinates with the predefined spatial position of the observer, the rays that can be observed are selected to
construct the received signal in the next step. Rays that do not reach the observer still provide meaningful insights. 
The observer plane, which is partitioned into discrete bins, contains the intersection point of each ray. 
The density of intersection points within each bin is computed, yielding a spatial distribution of light intensity.
This distribution corresponds to the concept of gain in lensing literature.

\subsection{Diffractive Regime}\label{Diffractive Section}
The physical essence of the diffractive regime resides in the computation of the Fresnel–Kirchhoff integral, which is discussed in detail by \citet{Narayan_1992}.
The received amplitude $\psi(X, Y)$ at a certain point $(X, Y)$  on the observer plane is calculated by integrating the contributions of light from every point $(x,y)$ on the screen, as:
\begin{equation}
    \psi(X,Y) = \frac{e^{-i\pi /2}}{2\pi r_{\rm F}^{2}}\int\int \exp \left[i\phi(x,y) + i \frac{(x-X)^{2}+(y-Y)^{2}}{2r^{2}_{\rm F}}\right]dxdy,
\end{equation}
where $r_{\rm F}$ is the Fresnel scale of this system\footnote{The Fresnel scale definition $r_{\rm F} = \sqrt{\lambda d/2\pi}$ varies across literature, regarding the presence or absence of the $2\pi$ factor in the denominator. While the effective propagation distance $d$ may adopt slightly different interpretations depending on the astrophysical context, it generally represents the effective distance of the system.}.  
The first term in the exponential represents the phase change contributed by the plasma screen, while the second term accounts for the geometric phase associated with the propagation of light.
In contrast to the refractive regime, the diffractive regime does not trace individual light rays but employs statistical description.
When the characteristic scale of plasma inhomogeneities is exceedingly small, resolving each bump with discrete grid points would require an impractically high resolution. 
This constraint necessitates the implementation of coarser computational grids with larger cell spacing to resolve macroscopic density variations in the plasma screen.
Within each grid cell, the effects of subgrid-scale variations are statistically described.

In this regime, the scattering screen manifests frequency-dependent deflective properties.
Light scattered within a critical angular range is detectable at the observer, while those beyond this threshold are geometrically excluded. 
This behavior is quantified through the deflection radius $R_{\rm deflect}$, which is defined as the region within which the screen provides sufficient deflection to redirect light toward the observer. 
The formalism assumes complete collection and integration of all radiation within $R_{\rm deflect}$, while neglecting contributions from exterior regions due to the inadequate deflection capability.

The numerical integration comprises two critical steps: (i) determination of the integration domain and (ii) evaluation of the phase contributions for each integration element. 
Following the thin screen formalism for scattering screen developed by \citet{Beniamini_Faraday_Depolarization}, the 3D scattering screen is compressed along the main propagation direction to form the equivalent  two-dimensional (2D) thinscreen. 
After compression, the screen is represented by its column density. 
The computational elements are rectangular patches bounded by four adjacent grid points, with the position of each patch represented by the center of the corresponding rectangle.
Although the theoretical integration domain is nominally defined by the deflection radius $R_{\rm deflect}$,  practical implementation requires more considerations. 
The effective radii $R_{\rm eff}$, which define the integration region, are determined based on several additional conditions. 
The finite beaming angle $\theta_{\rm FRB}$ creates a circular illumination zone on the plasma screen with radius: $R_{\rm FRB}  = d_{\rm sp} \theta_{\rm FRB}$, where $d_{\rm sp}$ is the source-to-screen distance. 
In addition, the screen has a physical boundary, $R_{\rm screen}$, beyond which no plasma is present and light propagates directly away without deflection, never reaching the observer. 
Another critical parameter is $R_{\rm deflect}$, which has been defined. 
The deflection radius $R_{\rm deflect}$ is frequency-dependent and can be estimated as $R_{\rm deflect} = d_{\rm sp}\theta_{\rm deflect}$, where $\theta_{\rm deflect}$ represents the maximum bending capability of the screen. 
The deflection angle, $\theta_{\rm deflect}$, is approximated as $N\theta_{\rm diff}$, where $N = l_{\rm Thickness}/l_{\rm \nu}$ is the number of scattering events and $\theta_{\rm diff} \sim \lambda / l_{\nu}$ represents the diffraction angle for a single scattering event. 
Here, $l_{\rm Thickness}$ denotes the thickness of the plasma screen, and $l_{\nu}$ is the characteristic effective bump size for each frequency. 
The parameter $l_{\nu}$ should be specified either as a free parameter or provided as additional input.
Therefore, $R_{\rm deflect}$ is given by:
\begin{equation}
    R_{\rm deflect} = d_{\rm sp}\theta_{\rm deflect} =  d_{\rm sp}N\theta_{\rm diff} = d_{\rm sp}\frac{l_{\rm Thickness}}{l_{\rm \nu}}\frac{\lambda}{l_{\rm \nu}} = d_{\rm sp}\frac{l_{\rm Thickness}c}{\nu l_{\rm \nu}^{2}}.
\end{equation}
This relation is only a rough estimate.
In more realistic scenarios, one must either adopt specific assumptions for the statistical treatment of scattering or simulate each scattering event at high resolution to obtain accurate results. 
For example, \citet{Beniamini_Faraday_Depolarization} modeled a nearly uniform screen with density fluctuations following a Kolmogorov turbulent cascade. 
Under these conditions, the relation becomes $R_{\rm deflect} \propto \nu^{-2.2}$. 
This relationship can be adjusted based on different assumptions or models,  and varying the model parameters changes the outcome. 
Finally, the effective radius $R_{\rm eff}$ is defined as the minimum among $R_{\rm FRB}$, $R_{\rm screen}$ and $R_{\rm deflect}$.

This computational method adheres to the conventional approaches of refractive regimes, ensuring systematic recording of light propagation times, phase shifts, and DM values. 
The key difference is that the screen is compressed, and the iteration occurs only once, implying a 2D thin screen model.

The large patches discretized from the 2D screen accommodate macro-scale variations, providing greater flexibility and facilitating the transition to the refractive regime.
In the code, 2D patches within the effective radii are considered, with each patch having only one incident ray at its center to enhance computational efficiency. 
The information carried by the light is then collected to construct the FRB signal in the next step.

\subsection{Supplementary Treatments for Both Regimes}
While the preceding sections detailed regime-specific core computational methodologies, this section presents additional shared processing steps. 
Although earlier analyses emphasized light–plasma interactions within the plasma screen, vacuum propagation segments crucially contribute to geometric components. 
For computational efficiency, the nonscreen regions are modeled as vacuum domains, intentionally neglecting uniform background plasma effects that can be conveniently incorporated through linear superposition. 
This hybrid treatment enables comprehensive parameter tracking throughout the full propagation path that combines vacuum and plasma segments.

Each light ray carries critical information along its propagation path, including the total travel time $t_{\rm total} = t_{\rm sp} + t_{\rm plasma} + t_{\rm po}$ (where ``s" stands for source, ``p" for plasma, and ``o" for observer), total phase shift $\phi_{\rm total} = \phi_{\rm sp} + \phi_{\rm plasma} + \phi_{\rm po}$, DM value, and additional parameters related to magnetized plasma (which are beyond the scope of this paper).  
The contributions of the plasma segment have already been discussed. Vacuum contributions, $t_{\rm sp}$ and $t_{\rm po}$ are computed based on the distances between the source and the incident point on the plasma screen, and between the exit point on the screen and the observer, respectively, divided by the speed of light in vacuum $c$. Besides, the wave number in the vacuum is given by $k = \omega/c$, which leads to the two geometric phase shifts: $\phi_{\rm sp} = \omega t_{\rm sp}$ and $\phi_{\rm po} = \omega t_{\rm po}$.

The total travel time $t_{\rm total}$, total phase shift $\phi_{\rm total}$, and accumulated DM value through the entire plasma form the dataset carried by the light rays and will be used to construct the FRB signal in the next process step.

\subsection{Light Reception Process} \label{ReceiveLight}
The synthesis of received signals occurs through the coherent superposition of light that reaches the observer. 
This critical process ensures the accurate encoding of astrophysical information in the signal. 
This section details the method of superposing the light to generate the final observed signal. 
The process begins with the handling of monochromatic light, after which signals from different frequencies are combined according to the resolution of the telescope.

\subsubsection{Monochromatic Signal Processing}
The Light Receiver module processes discrete-frequency light generated by the simulator through specialized signal construction routines. Because light of the same frequency can interfere and cause variations in intensity, meticulous treatment is required. 
This involves two steps: first grouping the light by frequency, then performing coherent summation of all components sharing the same frequency.

Within the computational framework, light is represented by complex electric field vectors. Any given electric field vector $\vec{E}$ could be decomposed into a linear combination of two orthonormal bases. For example, $\vec{E}$ can be written as:
\begin{equation}
    \vec{E} = E_{\rm 1}\hat{i} + E_{\rm 2}\hat{j},
\end{equation}
where $E_{\rm 1}$ and $E_{\rm 2}$ are the complex amplitudes along the $i$- and $j$-axes , which are perpendicular to the propagation direction 
$\hat{k}$, respectively. 
These amplitudes are given by:
\begin{equation}
    E_{\rm 1} = E\cos{\chi}e^{i\phi}, \quad E_{\rm 2} = E\sin{\chi}e^{i\phi}.
\end{equation}
Here, $\chi$ represents the polarization angle, $\phi$ denotes the phase, and $E$ is the scalar amplitude of the electric vector. 
Since this paper focuses on non-magnetized plasma, $\chi$, which is mainly associated with polarization information in magnetized scenarios, does not play a significant role in the current context. 
This representation format facilitates the subsequent addition.

For a given time bin, all rays arriving simultaneously should be summed as follows:
\begin{equation}
    E_{\rm 1,\text{total}} = \sum_{m} E_{{\rm 1},m}, \quad E_{\rm 2,\text{total}} = \sum_{m} E_{{\rm 2},m}.
\end{equation}
The received intensity at this frequency for the given time bin is then calculated as:
\begin{equation}
    I = E_{\rm 1,\text{total}} E_{\rm 1,\text{total}}^{*} + E_{\rm 2,\text{total}} E_{\rm 2,\text{total}}^{*}.
\end{equation}
This systematic treatment of monochromatic signals guarantees precise incorporation of interference phenomena, thereby leading to a robust construction of the observed light. 

\subsubsection{Superposition Across Multiple Frequencies}
Due to the finite spectral and temporal resolutions of observational instruments, rays with different frequencies and arrival times may be mapped to the same grid cell in the waterfall plot, a visual representation of signal intensity over time and frequency.
Consequently, the intensity of each grid cell must be computed by summing contributions from all constituent rays within it. 
Owing to the incoherent nature of polychromatic radiation, the Stokes parameters of these contributions in the same grid cell can be combined straightforwardly through linear addition.

For an intrinsic pulse, the observed lightcurve and spectral profile differ from the original signals due to plasma-induced multipath propagation, which introduces time delays across different trajectories. Thus, the observed signal morphology depends on two factors: plasma propagation effects and the intrinsic signal profile.

During the millisecond-scale duration of an FRB, the plasma distribution can be treated as static. 
Therefore, plasma-induced ray trajectories need not be recalculated throughout the burst.
Instead, they are computed once, stored, and then used by the receiver module to synthesize the observed signal as the intrinsic signal evolves. 
The intrinsic intensity evolution, characterized by the original lightcurve and spectrum $I_{\rm intrinsic}(\nu,t)$, is modified by plasma-induced time delays and phase shifts, producing the observed intensity distribution $I_{\rm received}(\nu,t)$.
At the receiver, the finite temporal and spectral resolutions of the instrument define discrete time-frequency grids, and all contributions within each grid cell are summed to construct the waterfall plot.

\section{Code Overview} \label{sec:Code}
Having discussed the theoretical method, this section introduces the technical aspects of the code. 
The code is made up of three main components: the Screen Generator, the Light Propagation Simulator, and the Light Receiver.
\begin{itemize}
    \item Screen Generator: Creates discretized screen grids based on specified density distributions.
    \item Light Propagation Simulator: Determines the path of light from the source through the plasma screen to the observer, while recording information along the way.
    \item Light Receiver: Constructs the received signal by superposing the simulated light based on the intrinsic lightcurves and spectra, along with the temporal and spectral resolutions of the telescope.
\end{itemize}

\subsection{Screen Generator}
The primary function of the screen generator is to create a plasma screen that accurately represents the desired electron density distribution, which is crucial for calculating light propagation through the plasma. 
This screen is described by discrete grid points generated under specific requirements.
Typically, a base distribution, such as a uniform or Gaussian function, serves as the foundation, onto which perturbations can be applied. 
Structures at various scales, such as clumps, can
be superimposed as well. 
Alternatively, density values can be directly imported from the results of other simulations. 
The screen generator then formats these data for compatibility with the light propagation simulator. 

\begin{figure}[ht!] 
\centering 
\includegraphics[width=1\textwidth]{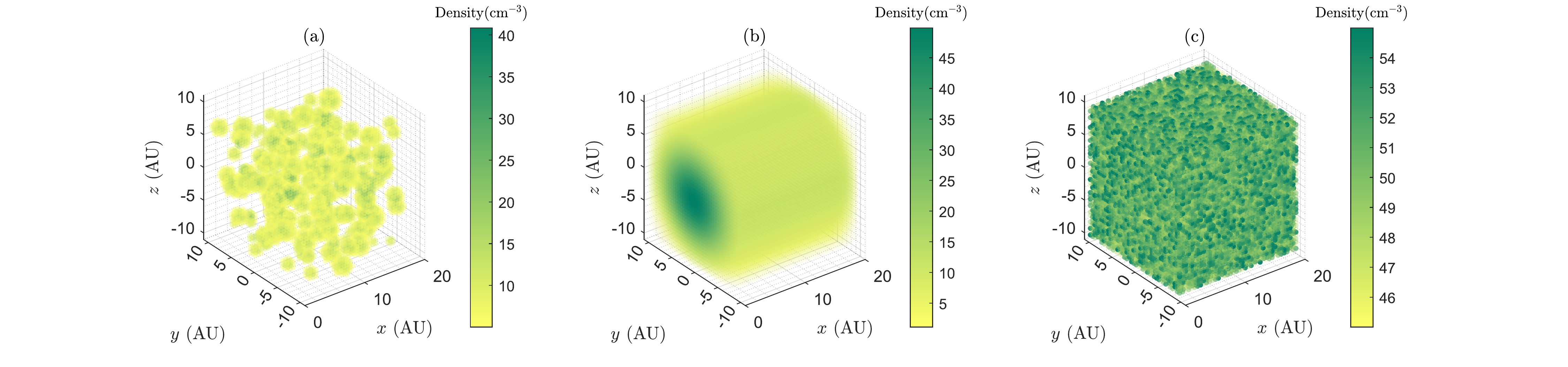} 
\caption{3D density distributions of the plasma screen. The color bar indicates plasma density, with transparency incorporated to visualize interior density; lower densities correspond to higher transparency. Panel (a) shows a screen with irregularly distributed 3D Gaussian blobs, sized from 0.5 au to 2.5 au in 0.5 au increments, with 50 blobs per size. Each blob is defined by a Gaussian function with a central density of 20 $\rm cm^{-3}$, fluctuating by $\sim$ 0.1, and randomly positioned with some overlap. Panel (b) shows a 2D Gaussian density distribution that remains constant along the x-axis, with a central density of 50 $\rm cm^-{3}$ and a cylinder width of 5 au. Panel (c) illustrates a scattering screen composed of small, randomly distributed clumps, with an average density of 50 $\rm cm^-3$ and random fluctuations of approximately 0.1. The fluctuations occur on the scale of the spatial resolution, corresponding to the grid point spacing of 0.5 au in this plot.}

\label{Fig:density_example} 
\end{figure}

Figure~\ref{Fig:density_example} demonstrates three representative types of 3D density distributions. 
Panel (a) displays a screen populated with irregularly distributed 3D Gaussian blobs. 
Each blob is well described by a 3D Gaussian function with a central density peak, and the overall distribution of blobs exhibits random variations in size and position.
However, the number of blobs is insufficient, meaning the distribution lacks statistical properties, which makes it
difficult to be treated analytically. 
Panel (b) shows a 2D Gaussian density distribution (with no variation along the $x$-axis), which can be explicitly defined by an analytical function. 
Panel (c) depicts a scattering screen composed of small, randomly distributed clumps. 
This screen features a uniform average density with random local fluctuations, emphasizing its statistical nature. 
The statistical properties of such a distribution can typically be described analytically, as is common in scattering screen scenarios. 
Each discrete grid point in the plasma screen permits individual adjustment, allowing precise control over local electron density.

\subsection{Light Propagation Simulator}\label{LightTripSimulator}
The light propagation simulator traces light trajectories from the source through the plasma screen to the observer plane.
During this process, the simulator tracks key information at each step, including time delays, phase shifts, and the number of electrons encountered, which are determined by the plasma properties.
\subsubsection{Basic Setup}
\begin{figure}[ht!]
\centering
\includegraphics[width=0.5\textwidth]{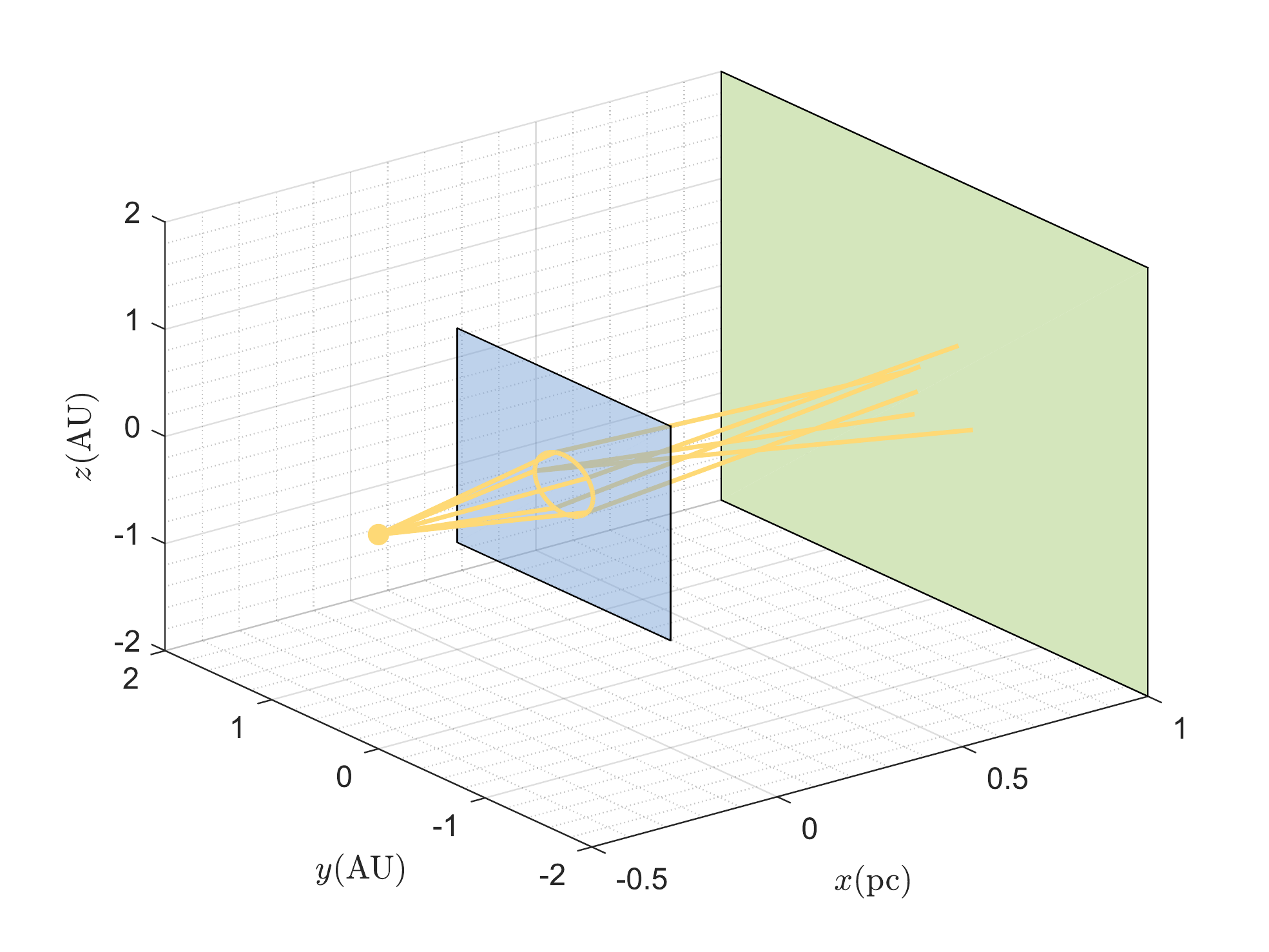}
\caption{Schematic of the system configuration. The source and observer are aligned along the x-axis, with the plasma screen positioned perpendicular to this axis, and its closest surface to the source located at $
x=0$. This setup illustrates the configuration for calculating light propagation through the plasma screen. The diagram is a simplified, exaggerated schematic for illustration and does not represent actual physical parameters. The screen has a thickness, but this representation is not to scale and is meant to provide a basic conceptual overview.}
\label{Fig:Sketch}
\end{figure}

A Cartesian coordinate system is established such that its $x$-axis aligns with the line connecting the source to the origin of the observer plane, with the observer plane positioned along the positive $x$-direction (Figure~\ref{Fig:Sketch}). 
The plasma screen is placed perpendicular to the $x$-axis, with its closest surface to the source located at $x = 0$.
The transverse displacement of the center of the plasma screen relative to the $x$-axis is quantified through two orthogonal displacement parameters: $\delta_{y}$ and $\delta_{\rm z}$, representing offsets along the $y$- and $z$- axes, respectively. 
The initial direction of light, represented by a normalized vector from the source to each incident point on the screen, is approximately $(1,0,0)$. 
This is because the distance $d_{\rm sp}$ from the source to plasma screen is considerably longer than the lateral dimensions of the screen.
The unique position of each incident point introduces only minor variations in the direction of light, leading to subtle directional discrepancies.

\subsubsection{Implementation Details}
To optimize computational efficiency, several strategies are employed: 
(a) Theoretically, light rays should be emitted randomly from the source within the half-opening angle, each intersecting the plasma screen at a unique incident point.
However, instead of simulating physical emission from the source, the implementation uses incident points to define the light.
This means the distribution of incident points is generated first, followed by tracing the full light trajectory backward to the source. 
(b) While Monte Carlo sampling is ideal for random incident point selection, a grid-aligned approach is adopted to improve computational efficiency and leverage known density distribution. 
Each rectangular region bounded by four neighboring grid points is treated as a computational unit. 
To avoid redundant calculations only one incident point per rectangular region is selected. 
If two incident points fall in the same rectangle, their calculations become nearly identical, rendering unnecessary computations redundant. 
At the same time, to adequately resolve plasma screen features, the grid resolution must be fully utilized. 
Each computational unit must include at least one incident point to reflect local density variations. 
Thus, the incident plane (the $y$–$z$ plane) is naturally divided into discrete patches, each defined by its center coordinates and size matching the grid resolution. 
The incident points (centers of patches) are uniformly distributed, with each patch acting as a computational unit. 
(c) Given the millisecond-scale duration of a burst, the density distribution of the screen could be considered static, and the trajectories do not change during the burst. 
Thus, repeated light emission simulations are unnecessary. 
Trajectories are precomputed once and stored. 
Observed signal variations arise solely from the intrinsic lightcurve and spectrum, modulated by stored plasma-induced delays and phase shifts.
The light receiver module synthesizes the observed signal using this stored trajectory pool, explaining why lightcurve and spectrum processing do not occur in the simulator. 
(d) While all plasma patches on the screen should ideally be considered, principles are applied to help reduce the number of patches.  
Only plasma-containing regions within the FRB beam half-opening angle $\theta_{\rm FRB}$ are considered. 
Patches located in nonilluminated regions, outside the radius $R_{\rm FRB} = d_{\rm sp}\tan{\theta_{\rm FRB}}$, or within plasma-free regions are removed.
In case the screen does not intersect the $x$-axis or has no deflection at the center, a special patch at (0,0) on the $y$–$z$ plane is always added. 
(e) Each patch is assigned a unique ID to track its light propagation path. 
This enables efficient retrieval of observable light data for signal construction in the light receiver module. 
(f) For the diffractive regime, patches within the effective radius $R_{\rm eff}$ are assumed to reach the observer. 
Although trajectory calculations are unnecessary here, the predefined trajectories must be traced once to obtain the parameters required for signal construction.

\subsection{Light Receiver}
The light receiver module processes light propagation data from the light propagation simulator module, following the methodology outlined in Section~\ref{ReceiveLight}. 
The input lightcurves and spectra are discrete functions defined over time and frequency, representing the intrinsic intensity $I_{\rm intrinsic}(\nu,t)$.
For each discrete time-frequency bin, the intrinsic intensity is adjusted by the time delay associated with its propagation trajectory, generating the modified intensity distribution $I_{\rm received}(\nu,t)$. 
By taking into account the temporal and spectral resolution of the telescope, contributions from all rays are integrated to synthesize the received waterfall plot. The intensity of a certain grid on the waterfall plot is calculated by summing all contributions within that grid:
\begin{equation} 
I_{\rm grid, ij} = \sum I_{\rm received}(\nu_{i-1}<\nu<\nu_{i}, t_{j-1}<t<t_{j}), 
\end{equation}
where $i$ indexes frequency grids defined by the spectral resolution of the telescope, and $j$ indexes time grids determined by its temporal resolution.

Additionally, for the refractive regime, further processing is applied to determine the nonuniform light distribution on the observer plane caused by plasma effects. 
The light propagation simulator module outputs intersection points on the observer plane as 2D coordinates, where each point is defined by two values (e.g., (1, 2)). These points are plotted on a single plane, which is partitioned into small bins. 
The count of points within each bin is normalized relative to a reference bin unaffected by plasma lensing. 
This procedure produces the 2D light distribution on the observer plane.

\section{Results} \label{sec:Results}
This section presents numerical results, organized into two parts. 
The first part focuses on code validation by comparing numerical outcomes with analytical solutions. 
The second part explores scenarios beyond the scope of analytical descriptions, demonstrating the capabilities of the code in handling complex situations. 
Throughout the section, the input lightcurve and spectrum remain fixed. 
A rectangular pulse with a duration of $2$ ms is used for the lightcurve and a flat-top rectangular spectrum spans $1-1.5$ GHz.

\subsection{Validation of Standard Results}
This section rigorously evaluates the capabilities and convergence of this code. 
Its performance is validated through systematic comparison of computational results against analytical solutions across multiple benchmarks. 
First, the ability of this code to reproduce four nonmagnetized plasma-induced effects—dispersion, plasma lensing, scintillation, and scattering—is demonstrated, establishing a baseline for its standard outputs.
Subsequently, numerical convergence is assessed by incrementally increasing the grid resolution to validate its reliability.

\subsubsection{Dispersion Test}
\begin{figure}[ht!]
\centering
\includegraphics[width=1\textwidth]{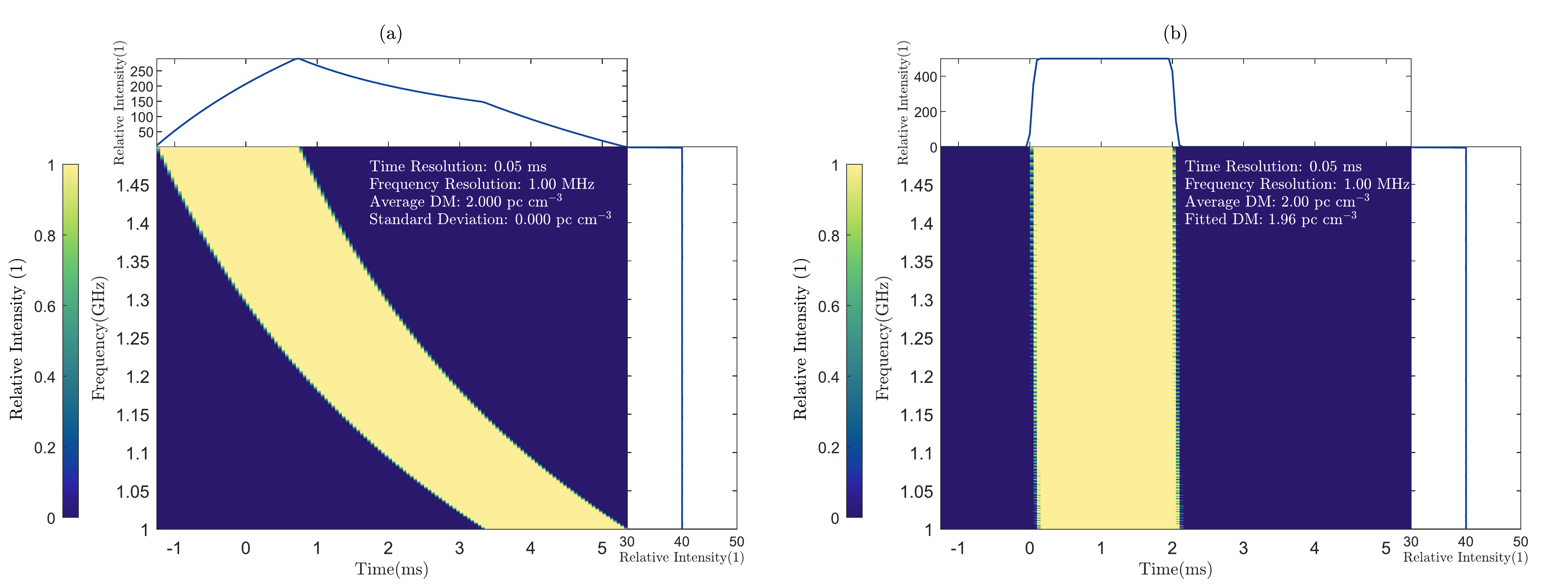}
\caption{Simulated dispersed signal before (left) and after (right) dedispersion. The left panel is the waterfall plot of a dispersed signal with DM of $2\ {\rm pc\ cm^{-3}}$, showing a distinct diagonal pattern (top-left to bottom-right) caused by frequency-dependent delays. Lower frequencies experience longer delays. The right panel displays the dedispersed waterfall plot, processed with a fitted DM of $1.96\ \ {\rm pc \ cm^{-3}}$. The dispersion-corrected signal approximates a flat-top square wave, with residual edge distortions due to finite temporal resolution and dedispersion inaccuracies. }
\label{Fig:WaterFalls}
\end{figure}

The dispersion effect, a ubiquitous phenomenon in the presence of plasma, manifests as the time delay between the low- and high-frequency components of the signal. 
This delay, termed dispersion delay, is defined as
\begin{equation} \label{dispersion_equation}
\Delta t = t(\nu_{1}) - t(\nu_{2}) = \frac{e^{2}}{2\pi m_{\rm e} c} \left( \frac{1}{\nu_1^2} - \frac{1}{\nu_2^2} \right) \rm{DM} \simeq (4.15\ \rm{ms}) \left( \frac{1}{\nu_{{1}, {\rm GHz}}^2} - \frac{1}{\nu_{{2}, {\rm GHz}}^2} \right) \frac{\rm{DM}}{\rm{pc\ cm}^{-3}}, \end{equation} 
where $\nu_2 > \nu_1$ and the condition $\omega \gg \omega_p$ holds \citep{zhang_physics_2023}. 
To isolate and validate the dispersion effect in the code, other plasma-induced phenomena (lensing, scintillation, scattering) are excluded. 
The light is set to propagate through a uniform plasma screen with a density of $20\ {\rm cm^{-3}}$ and a thickness of $0.1\ {\rm pc}$. 
Since no deflection occurs, only light traveling along the $x$-axis (from the source to the observer) is received.
Simulated results are shown in panel (a) of Figure~\ref{Fig:WaterFalls}, displaying the raw signal waterfall plot without dedispersion. 
This plot acts as a reference for the standard output of this code, annotated with the true DM value and the resolutions of the telescope. 
The true DM (statistical average of individual DM values from all rays reaching the observer) and its standard deviation are derived directly from the ray-tracing results. 
Each received ray contributes a unique DM value during superposition at the observer.

The simulated signals achieve high fidelity to real observational data, thereby permitting application of analogous processing methodologies to extract physical insights.  
For instance, the dispersed signal shown in the left panel of Figure~\ref{Fig:WaterFalls}, can be dedispersed to estimate a fitted DM value. 
The dedispersed result, displayed in the right panel of Figure~\ref{Fig:WaterFalls}, demonstrates the recovery of a near-flat-top lightcurve. 
The optimal DM is determined by iteratively shifting the signal across trial DM values and identifying the value that maximizes the peak flux.\footnote{The dedispersion method iterates over a grid of trial DM values spanning the plausible range of DM values. For each trial DM value, frequency-dependent time delays are calculated using Eq.~\ref{dispersion_equation} and applied inversely to align frequency channels. The aligned data are integrated to generate trial lightcurves, with the optimal DM identified as the value maximizing the peak flux.  This approach is consistent with the incoherent dedispersion method described in the review by \citet{Petroff2019}.}  This yields a fitted DM of $1.96\ \ {\rm pc \ cm^{-3}}$, consistent with the true physical value of $2.00\ \ {\rm pc \ cm^{-3}}$ (relative error: $2\%$). 
While the dedispersed lightcurve approximates a rectangular pulse, its edges exhibit slight curvature rather than sharp transitions. 
This imperfection stems from finite temporal resolution: edge time bins capture incomplete signal segments, further exacerbated by residual inaccuracies in the dedispersion process.

\subsubsection{Plasma Lensing Test}
Plasma lensing refers to the phenomenon in which electron
density gradients in plasma structures deflect electromagnetic waves, effectively acting as lenses. 
As waves propagate through inhomogeneous plasma, they accumulate position-dependent phase delays, as described by Eq.~\ref{PhaseShift}, leading to deflections. 
This redistributes light on the observer plane, forming regions of enhanced or diminished intensity.
Observationally, plasma lensing modulates signal properties, inducing intensity fluctuations \citep{Clegg_1998}, time delays \citep{Er_2020}, and even the change of spectrum\citep{Cordes_2017, Kumar_2024_spectral}. 

Computationally, plasma lensing is modeled within the refractive regime through iterative recalculation of light-ray deflection angles at each propagation step. 
To validate this implementation, the following test evaluates the accuracy of predicted light deflection.

\begin{figure}[ht!]
\centering
\includegraphics[width=1\textwidth]{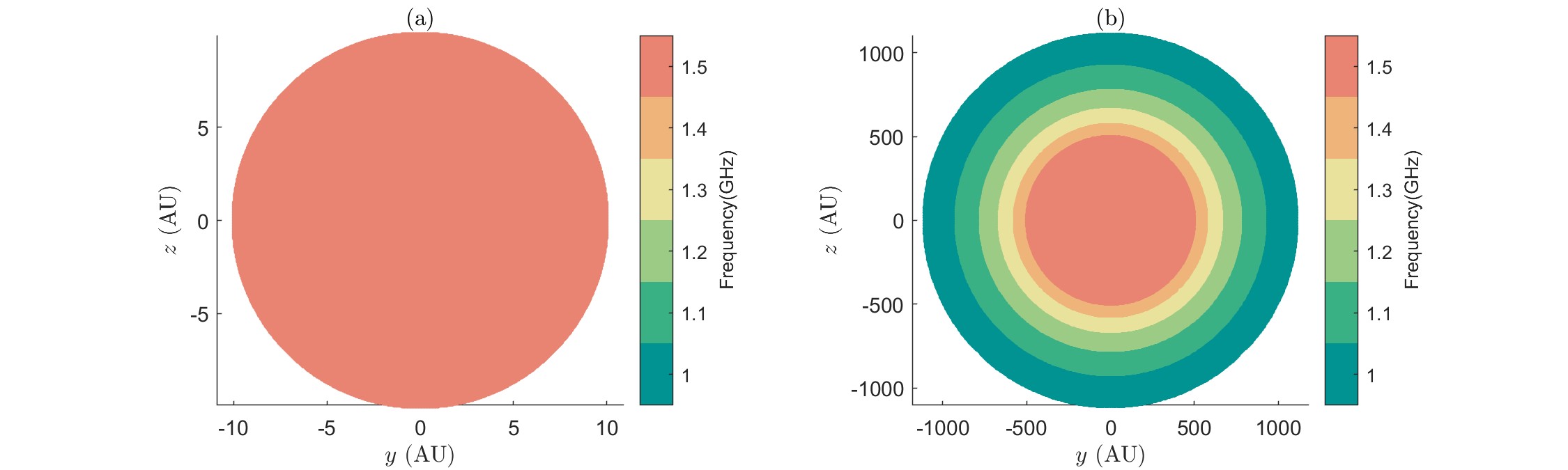}
\caption{Intersection points on the observer plane for distinct plasma configurations. Panel (a) demonstrates a uniform plasma configuration. Panel (b) depicts a 2D Gaussian plasma configuration (constant along the $x$-axis) with peak density $\rm 0.15 \ cm^{-3}$, thickness $0.1$ pc, and lens width $2$ au. Both configurations employ screens positioned $100$ kpc from the observer, with an infinite source and a 10 au circular illuminated region. Color mapping corresponds to frequency values as shown in the color bar.}
\label{Fig:ObserverPoints}
\end{figure}

After constructing the plasma density grid, the light propagation simulator computes light-ray trajectories through the medium (see Sections~\ref{Refractive Section} and~\ref{LightTripSimulator}), resolving deflection angles, time delays, and phase shifts. 
Although full trajectory reconstruction is impractical for explicit visualization, intersection points at the observer plane are captured via parametric coordinate mapping. 
Figure~\ref{Fig:ObserverPoints} illustrates typical arrival point distributions. 
Panel (a) corresponds to a uniform plasma screen, where no deflection occurs.
Intersection points are frequency independent and uniformly distribute across the observer plane. 
Because of the complete spatial overlap across frequencies, lower-frequency components become invisible in the visualization (only the highest frequency remaining visible). 
However, this is purely a graphical limitation--no physical obscuration occurs and observational data retains full spectral information. Incident parallel light that is initially confined to a $10$ au radius circular region on the plasma screen. 
Since the incident light is parallel, the observer plane also has a circle of $10$ au radius on the observer plane for the uniform case (panel (a)).
In contrast, panel (b) depicts a divergent plasma lens. 
Lower frequencies experience stronger deflections, spreading light over a broader area. 
The light disperses across $\sim1000$ au to $\sim1000$ au,
starkly contrasting with the uniform screen and confirming the lensing divergence.

\begin{figure}[ht!]
\centering
\includegraphics[width=1\textwidth]{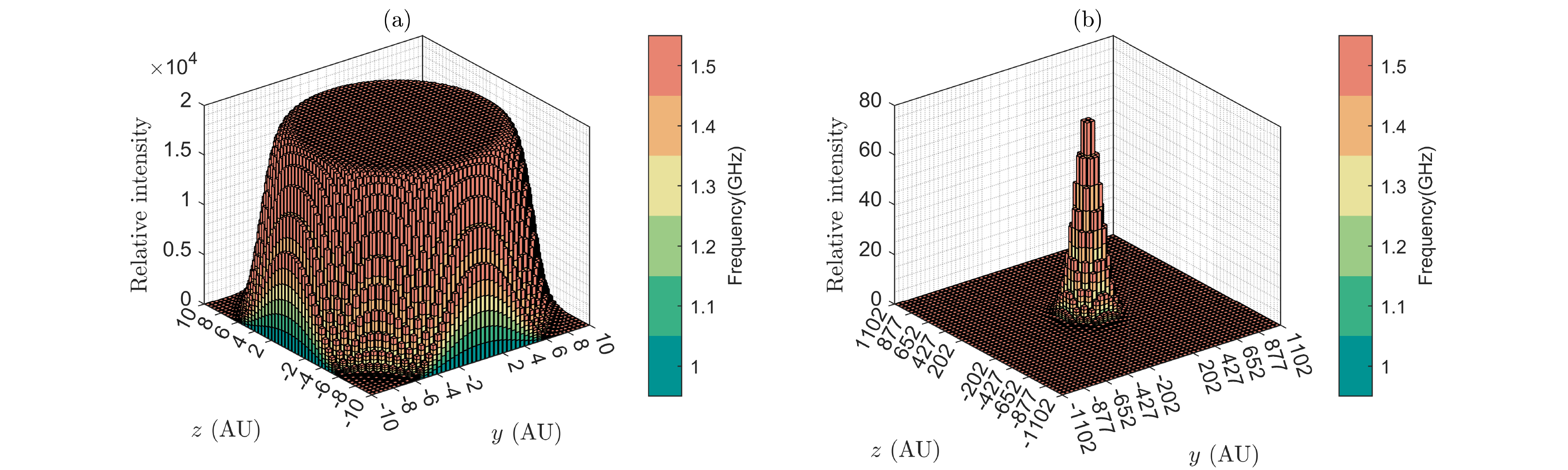}
\caption{Relative intensity distribution on the observer plane. Bar height represents relative intensity, calculated by binning intersection points and normalized to $10^6$ emitted rays per frequency. Panel (a) shows the uniform plasma case, where no deflection occurs, resulting in a uniform, frequency-independent distribution of light. The edge attenuation arises from numerical limitations in sampling density. Panel (b) presents the plasma lens case, where deflection induced divergence produces extended intensity distribution with caustic concentrations indicating divergent effects caused by plasma-induced deflection.}
\label{Fig:GainOnObserverPlane}
\end{figure}

By dividing the observer plane into discrete bins and statistically analyzing the density of intersection points within each bin, the relative gain induced by the plasma lens can be quantified. 
As illustrated in Figure~\ref{Fig:GainOnObserverPlane}, regions with higher intersection densities correspond to amplified intensity levels.
Panels (a) and (b) of Figure~\ref{Fig:GainOnObserverPlane} correspond to their counterparts in Figure~\ref{Fig:ObserverPoints}, representing uniform and divergent plasma configurations, respectively. 
Panel (a) represents the uniform plasma case where no deflection occurs, resulting in uniform gain across the observer plane. 
The apparent intensity decline at the edges is an artifact of incomplete sampling in peripheral bins, where fewer rays contribute. 
By contrast, panel (b) shows the divergent lens case that redistributes light nonuniformly, broadening the spatial spread through chromatic deflections. 
The divergent lens produces a frequency-dependent intensity distribution, concentrating light into caustic structures while depleting flux in the central region.
Caustic positions shift systematically with frequency, reflecting the chromatic nature of plasma lensing\footnote{Panel (b) may appear to exhibit light concentration; however, this is an artifact of the differing $y$- and $z$-axis scales between panels. The apparent bright regions arise from caustics formed by deflected light, rather than the convergent lens effect.}.

\begin{figure}[ht!]
\centering
\includegraphics[width=1\textwidth]{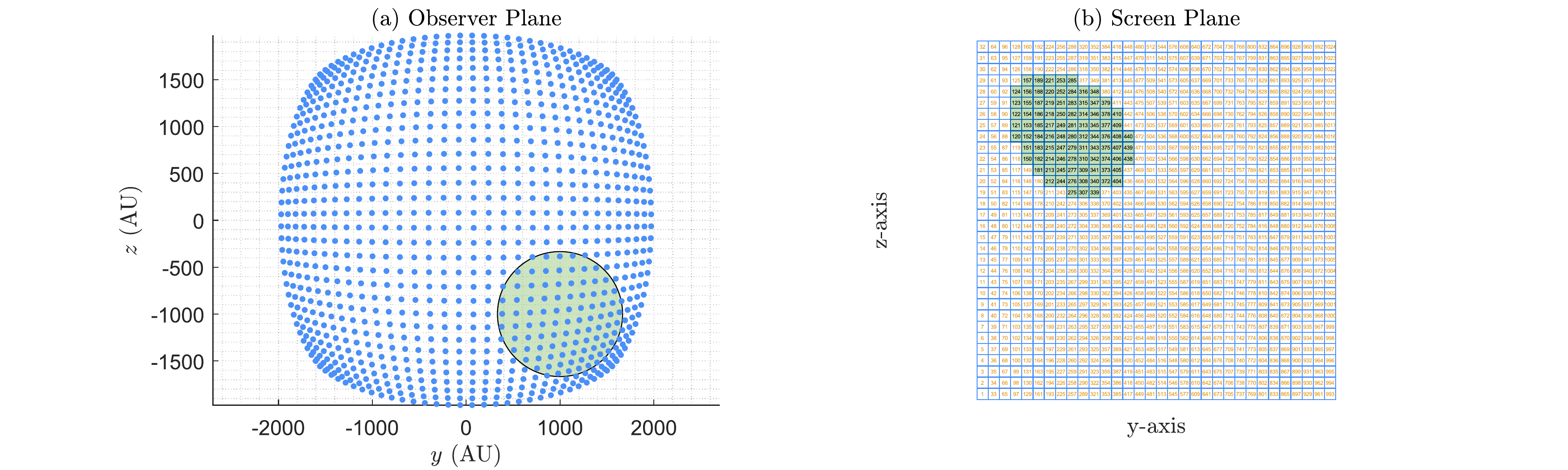}
\caption{Schematic for localization of targeted area. (a) shows the observer plane, with blue dots marking light arrival coordinates at a specific frequency, and the green region indicating the area of interest. (b) depicts the screen plane, which is divided into patches, where the green area corresponds to that in (a) and will be subdivided into finer grids to achieve higher resolution.}
\label{Fig:InterestArea}
\end{figure}

To analyze the received signal at a specific observer plane location, one may tally light rays across all frequencies arriving at that point. 
However, finite numerical resolution may result in few or no rays reaching the selected coordinate (see Section~\ref{ResolutionTest} for detailed discussion).
Global grid refinement across the entire screen is computationally wasteful for single point analyses. 
Instead, the region of interest and its immediate vicinity can be targeted, as demonstrated in  Figure~\ref{Fig:InterestArea} (a). 
By tracing individual light trajectories, the code identifies the specific screen patches passed through by each ray. 
This automated localization of relevant screen regions  (as shown in Figure~\ref{Fig:InterestArea} (b)), enables selective subdivision of the targeted area into high-resolution fine grids.

After collecting the light at a specific point, the next step is synthesizing the received signal by coherently superposing the light in the light receiver module, which generates the final FRB waterfall plot.

As the computation of light-ray deflection constitutes the core algorithm of the code, its numerical results necessitates rigorous validation through quantitative benchmarking. 
The deflection calculations are verified against analytical solutions for plasma lensing in prior theoretical studies. 
Previous works, including studies by \citep{Clegg_1998} and \citep{Cordes_2017}, model Gaussian plasma lenses as one-dimensional (1D) systems. To align with these models, the plasma screen is configured with no variation along the $x$-axis and $z$-axis, effectively simplifying the scenario to a 1D case. 
This enables direct comparison of the 1D numerical results with published analytical predictions.

 In a 1D plasma lens system, the fundamental quantity is the electron column density distribution, which, for a Gaussian profile, is expressed as
\begin{equation}
    N_{\rm e}(y) = N_{\rm 0}{\exp}\left[-(y/a)^{2}\right],
\end{equation}
where the column density $N_{\rm e} = n_{\rm e}s$ quantifies the integrated electron density along the line of sight through the lens, with $s$ denoting the path length, and $a$ representing the characteristic size of the lens.

The lensing effects depend not only on the plasma distribution but also on additional physical parameters.
Crucially, parameter degeneracy arises when distinct combinations of these inputs yield observationally indistinguishable lensing signatures. 
To facilitate quantitative characterization of the lensing effects, a dimensionless parameter $\alpha$ is introduced, which is recognized in current research as a fundamental descriptor of the lens characteristics. This critical parameter is defined \citep{Clegg_1998} as
\begin{equation}
    \alpha \equiv \lambda^{2}r_{\rm e}N_{\rm 0}D/\pi a^{2} = \left(\frac{\sqrt{\lambda D}}{a}\right)^{2}\frac{1}{\pi}\lambda r_{\rm e} N_{\rm 0},
\end{equation}
where $r_{\rm e}$ corresponds to the classical electron radius, $D$ specifies the distance between the plasma lens (for the case of an infinitely distant point source) and the observer. 

\begin{figure}[ht!]
\centering
\includegraphics[width=1\textwidth]{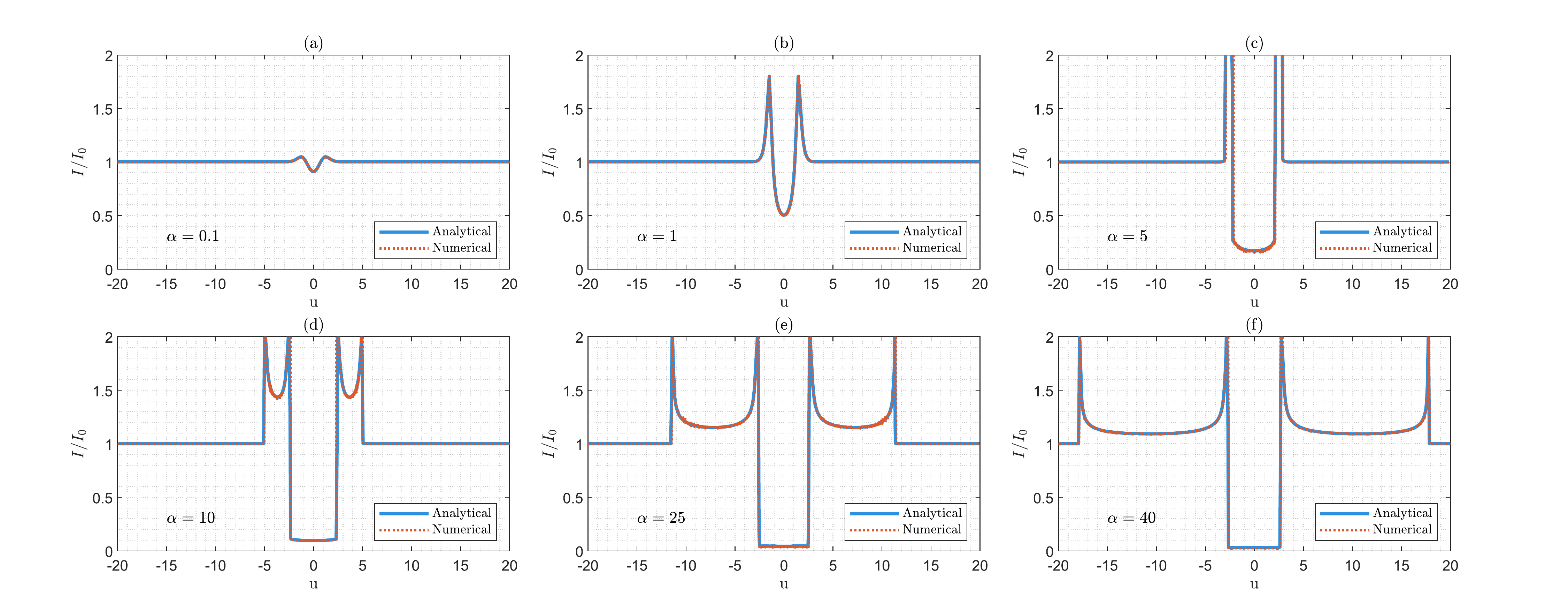}
\caption{Comparison of analytical (blue solid line) and numerical (red dotted line) results for light distribution on the observer plane across varying $\alpha$ values. Vertical axis displays normalized intensity $I/I_{\rm 0}$, where $I_{\rm 0}$ denotes unlensed uniform intensity. Horizontal axis shows the dimensionless transverse coordinate $u = y_{\rm obs}/a$, scaled by the characteristic lens size $a$.}
\label{Fig:Comparation}
\end{figure}

To validate the methodology, numerical parameters are calibrated to correspond with the configuration in \citet{Clegg_1998}, with comparative results presented in Figure~\ref{Fig:Comparation}.
The dimensionless intensity $I/I_{\rm 0}$ (where $I_{\rm 0}$ denotes unlensed uniform intensity) is plotted against the normalized transverse coordinate $u=y_{\rm obs}/a$, parameterizing positions on the observer plane. 
Numerical results (red dotted curves) demonstrate excellent agreement with analytical solutions (blue solid curves), exhibiting only minor numerical fluctuations. 
Quantitative error analysis employs grid-matched interpolation to compute pointwise relative error $|(I_{\rm num}- I_{\rm anal})/ I_{\rm anal}|$.
Mean relative errors across the profiles measure $0.3\%$, $0.5\%$, $9.2\%$, $10.3\%$, $3.9\%$ and $7.3\%$ for panels (a)-(f) respectively. 
Elevated errors in panels (c)–(f) originate from caustic regions, where analytical solutions diverge at singularities. Since numerical values cannot technically be divergent, significant errors occur near the critical divergent points. 
Crucially as shown in the figure, the numerical results maintain consistent with the analytical ones.

\subsubsection{Scintillation Test}\label{Scintillation}
Scintillation constitutes a fundamental plasma propagation effect in nonmagnetized media, characterized by the variations of intensities across temporal and spectral domains. 
This phenomenon originates from coherent electromagnetic wave interference through multipath propagation induced by plasma density fluctuations.
As radiation traverses inhomogeneous plasma structures, different phase accumulation along distinct paths generates interference patterns when light components reach the observer. 
In observations, scintillation manifests as characteristic intensity modulation patterns in dynamic spectra, exhibiting alternating bright and dark stripes in the frequency-time space.

In this framework, scintillation is inherently associated with the light receiver module, and its accurate reproduction validates the numerical implementation of coherent signal summation--the core functionality enabling interference modeling. 
Given the intrinsic coherence of FRB signals, interference effects demand rigorous treatment via the phased summation methodology detailed in Section~ \ref{ReceiveLight}. 
The phase difference, which determines whether interference leads to amplitude enhancement or cancellation, arises from both geometric path variations and dispersive effects due to electron density fluctuations (i.e., DM variations along the paths).
Although conventional scintillation theories predominantly address geometric phase contributions, this code self-consistently incorporates both geometric and dispersive phase components. 
To verify that the code properly handles phase differences and coherent signal superposition, an example considering only geometric phase shifts is provided. 
In this configuration, two patches (or slits) allow light from an infinitely distant source to pass through.  
One patch lies along the source-observer axis, and the other is parallel to the first but laterally offset by a distance $x$ within the plasma plane. 
Both patches are located at a distance $D$ from the observer plane. 
This arrangement mirrors the classic two-slit interference experiment, yet diverges fundamentally in its observational manifestation: while the traditional two-slit experiment focuses on the formation of bright and dark fringes at a fixed frequency on the observer plane, scintillation refers to the phenomenon in which an observer at a fixed location receives light spanning different frequencies. 
This configuration generates frequency-dependent interference through the geometric phase delay:
\begin{equation}
\Delta\phi  = \omega\Delta t = \omega\frac{\sqrt{x^{2}+D^{2}}-D}{c} \approx \omega\frac{x^{2}}{2Dc}. 
\end{equation}
Consequently, the phase differences between the light from the two patches vary with frequency, leading to constructive interference at some frequencies and destructive interference at others, thereby producing scintillation patterns that appear as stripes in the waterfall plot of the received signal.
One direct approach to estimate the scintillation bandwidth is formulated below, where the difference of phase differences between different frequencies can be expressed as
\begin{equation} 
\Delta(\Delta \phi) = \Delta t \Delta \omega \approx \frac{\pi x^{2} \Delta\nu}{Dc}. 
\end{equation}
To determine the scintillation bandwidth, we set 
$\Delta(\Delta\phi)  = \pi$ to obtain a noticeable change, leading to:
\begin{equation} \label{scintillation bandwidth}
\Delta \nu \sim \frac{Dc}{x^{2}}. 
\end{equation}

\begin{figure}[ht!]
\centering
\includegraphics[width=1\textwidth]{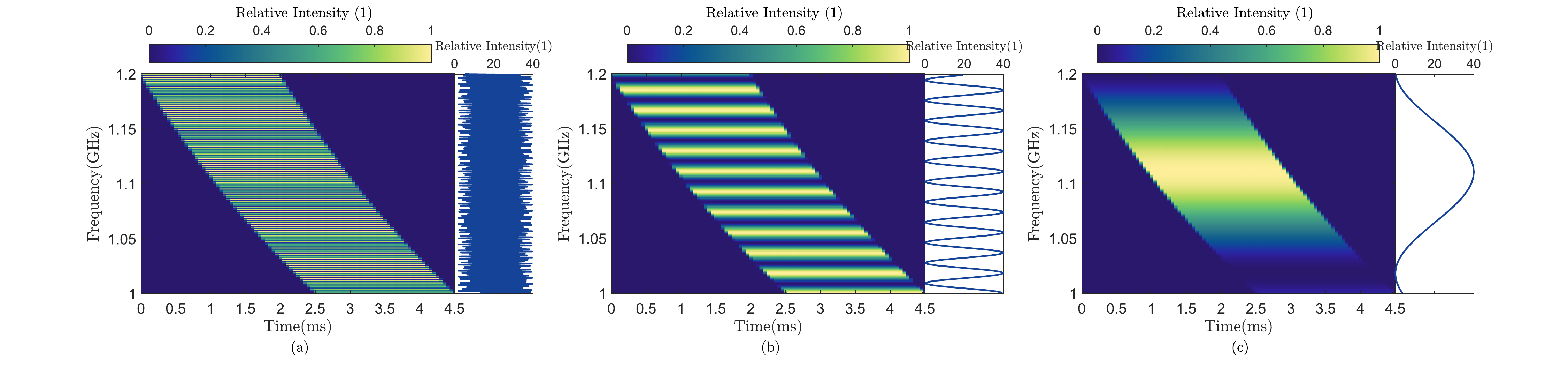}
\caption{Scintillation examples originate from two-path interference configuration. Panels (a) - (c) show geometrically distinct cases with plasma screen-observer distance $D = 1, 10, 100\ \rm kpc$. Light from an infinitely distant source passes through two plasma patches, one aligned with the source-observer axis, another transversely offset by $x = 10^{13}$ cm. Fourier transformation reveals scintillation bandwidths equal to $1.1$ MHz, $9.1$ MHz, and $100$ MHz respectively. The intrinsic signal is a $2$ ms rectangular pulse with frequency spanning $1-1.2$ GHz}
\label{Fig:Interference}
\end{figure}

Varied values of $D$ are applied in the panels of Figure~\ref{Fig:Interference} to demonstrate the ability of the code to effectively handle phase differences and coherent signal superposition.
In this setup, we fix $x=10^{13}$ cm and choose distances $D$ of $1$ kpc, $10$ kpc, and $100$ kpc for Panels (a), (b), and (c), respectively. 
The estimated scintillation bandwidths from Eq.~\ref{scintillation bandwidth} are $0.93$ MHz, $9.3$ MHz, and $93$ MHz. 
Fourier transformation yields corresponding bandwidths of $1.1$ MHz, $9.1$ MHz, and $100$ MHz, which align well with the analytical predictions, with relative errors of $18\%$, $2\%$, and $7.5\%$, respectively.
The elevated errors in cases (a) and (c) can be attributed to specific factors. When implementing a $1$ MHz spectral resolution, case (a) exhibits pronounced beat patterns caused by interference between the instrument resolution and the scintillation bandwidth. 
To mitigate this effect, panel (a) employs a spectral resolution of $0.5$ MHz, whereas Panels (b) and (c) maintain a resolution of $1$ MHz. 
Nonetheless, residual beat modulation persists, remaining higher relative errors in panel (a). 
In panel (c), the approximate $100$ MHz scintillation bandwidth occupies half of the $200$ MHz observational bandwidth, limiting the effective sampling window for Fourier transformation, which further contributes to the elevated relative error.

\subsubsection{Scattering Test}\label{Scattering}
\begin{figure}[ht!]
\centering
\includegraphics[width=0.9\textwidth]{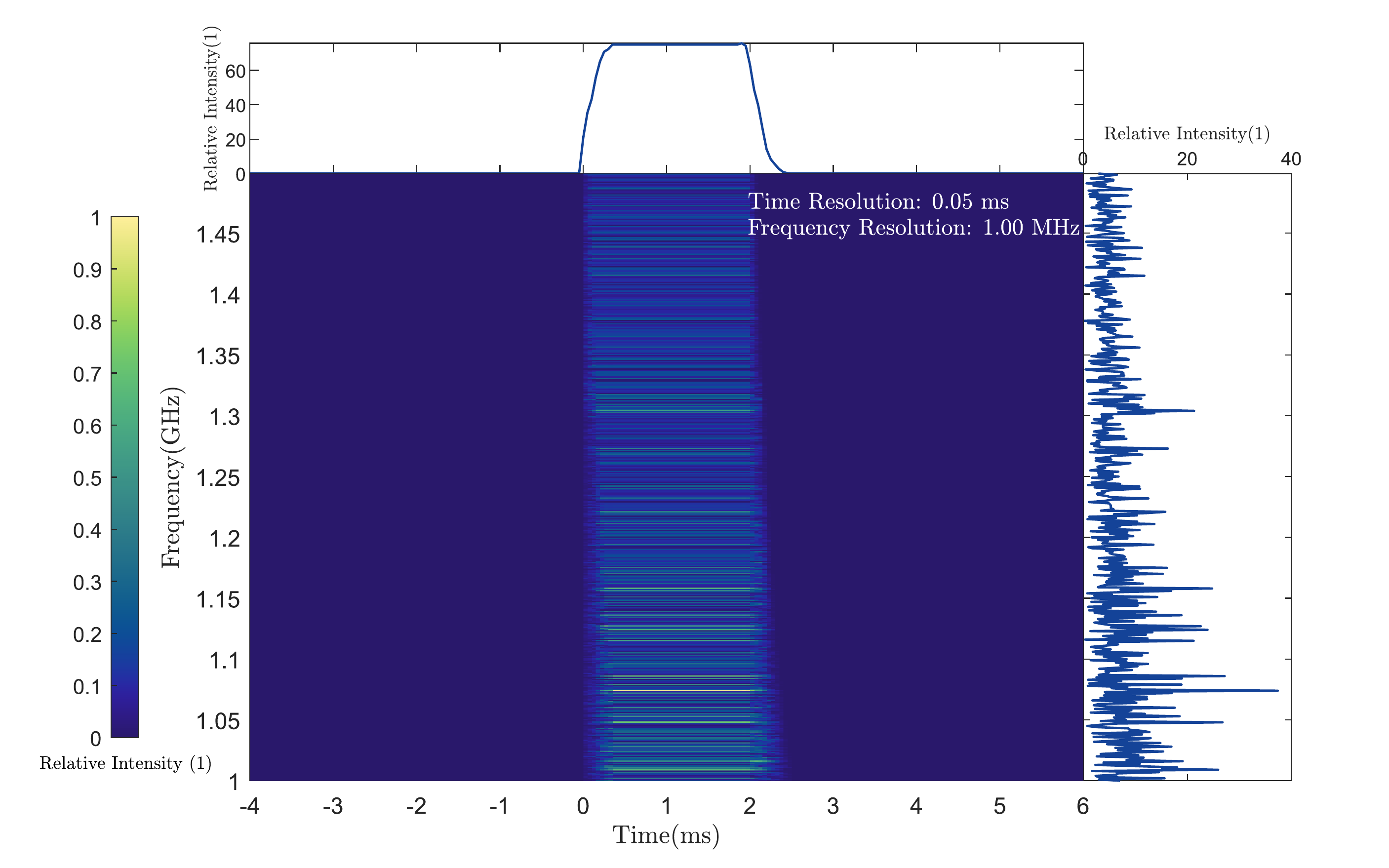}
\caption{Waterfall plot demonstrating frequency-dependent pulse broadening from plasma scattering.}
\label{Fig:ScatteringTail}
\end{figure}

Scattering, the final effect of nonmagnetized plasma examined in this study, manifests primarily as temporally broadened signals broadening, exhibiting strong frequency dependence with enhanced broadening at low frequencies. 
In the literature, this phenomenon is attributed to wave diffraction at a plasma scattering screen, which is implemented in the EMPI code through diffractive regime modeling. 
Only rays within a specific effective radius of the scattering screen can reach the observer.
Because these rays follow varying geometric trajectories from different points within the effective circle, their arrival times differ, resulting in an extended signal duration that appears as a scattering tail in the waterfall plot. 
At lower frequencies, the effective radius increases, leading to a wider range of arrival times, and consequently, more pronounced signal broadening. 
This characteristic tail-like feature remains evident even after dedispersion, as illustrated in Figure~\ref{Fig:ScatteringTail}.  (The figure also reveals a rising wing preceding the main pulse flattening, which comes from the intrinsic pulse being a top-flat square wave, deliberately chosen to accentuate these effects, with both the scattering tail and the rising wing being influenced by the dedispersion process).

\begin{figure}[ht!]
\centering
\includegraphics[width=0.5\textwidth]{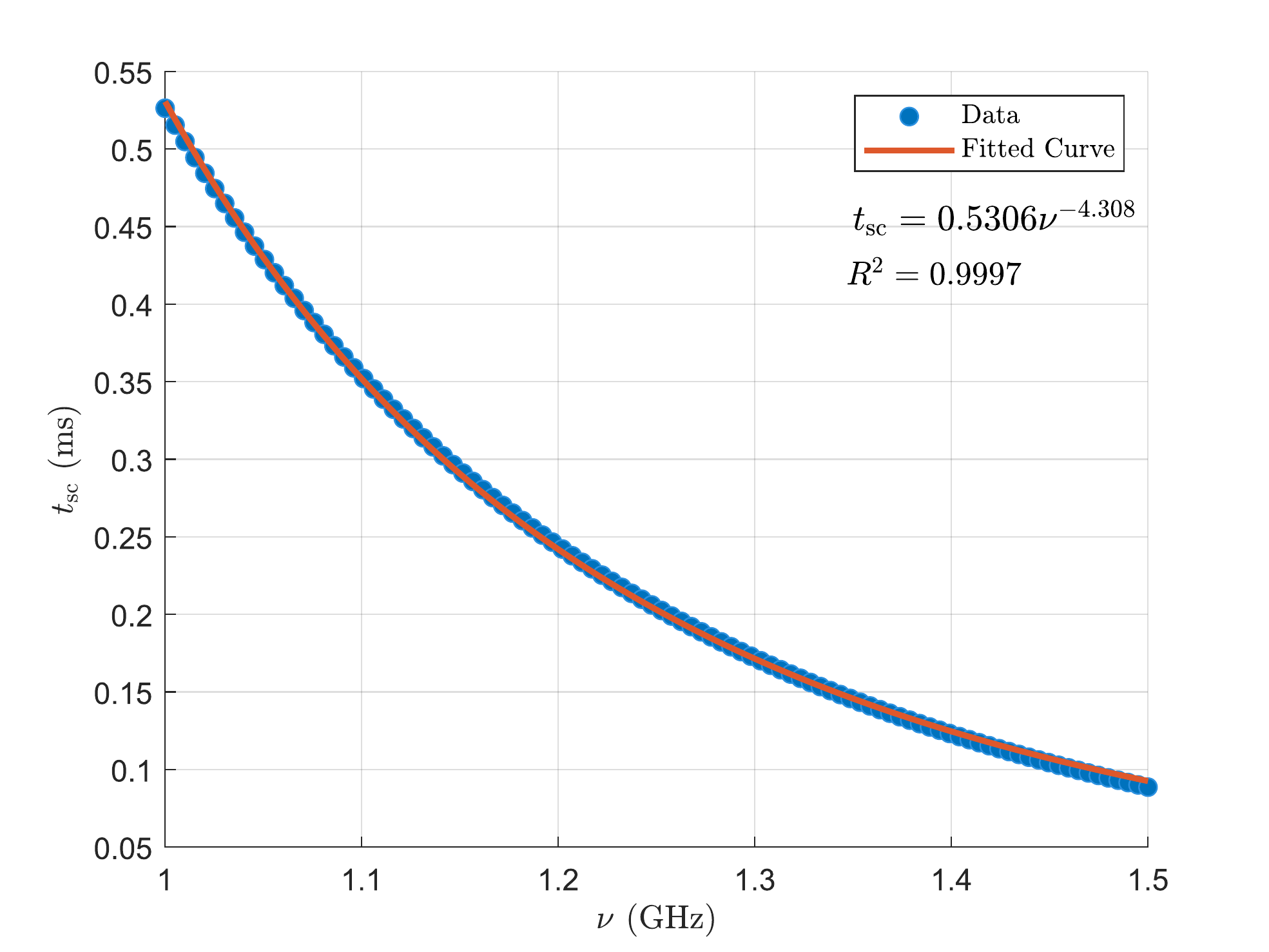}
\caption{Fitting of the scattering time scale as a function of frequency. Blue dots indicate the numerical data, while the red line represents the fitted curve. The best-fit relation is $ t_{\rm sc} = 0.5306\nu^{-4.308}$, with a correlation coefficient of 0.9997.}
\label{Fig:ScatteringFit}
\end{figure}

The spatial scales of plasma density variations across the scattering screen are typically much smaller than the thickness of the screen itself. 
As rays traverse the screen, the cumulative effect of these small-scale variations averages out, resulting in an approximately uniform plasma distribution.
Therefore, the time delay introduced by the plasma can be neglected, with differences in geometric path lengths dominating the time broadening.
Moreover, as described in Section~\ref{Diffractive Section}, the effective radius is frequency-dependent, with lower frequencies corresponding to larger effective radii and hence greater broadening.
In the test, a relation for the deflection ability of the screen is provided, $R_{\rm deflect} \propto \nu^{-2.2}$ \citep{Beniamini_Faraday_Depolarization}, assuming a point source located at infinity and a screen positioned at a distance $d = 100$ pc from the observer. 
For a point on the screen that deviates from the $x$-axis by a distance $R_{\rm deflect}$, the additional path length relative to the central point is given by
\begin{equation}
    \Delta = \sqrt{d^{2}+R_{\rm deflect}^{2}} - d \approx \frac{R_{\rm deflect}^{2}}{2d}.
\end{equation}
Under the exaggerated screen parameters used to demonstrate the scattering effects, at a frequency of $\sim 1$ GHz, the corresponding $R_{\rm deflect}$ is on the order of $\sim 10^{14}$ cm, yielding an estimated scattering time scale of about $0.5$ ms. 
The numerical results, presented in Figure~\ref{Fig:ScatteringFit}, indicate a fitted relation $t_{\rm sc} = 0.5306\nu^{-4.308}$ with a correlation coefficient of 0.9997. This result is consistent with the theoretical expectation $t_{\rm sc} \propto \nu^{-4.4}$, with a relative error of $2\%$.

\subsubsection{Convergence Test}\label{ResolutionTest}
For completeness, systematic resolution tests across key output parameters are presented in this section to validate numerical convergence. 
As with any grid-based calculation, discretizing the domain introduces truncation errors that diminish with resolution refinement.
While higher resolution generally improves accuracy by approaching the continuum solution, the convergence rate and threshold depend critically on system-specific physics.

The code under investigation produces two types of outputs -- the simulated received signal and the plasma lensing gain -- both of which are influenced by the resolution of the screen. 
The following demonstrates how these outputs converge to stable values as the resolution is progressively refined.

\begin{figure}[ht!]
\centering
\includegraphics[width=1\textwidth]{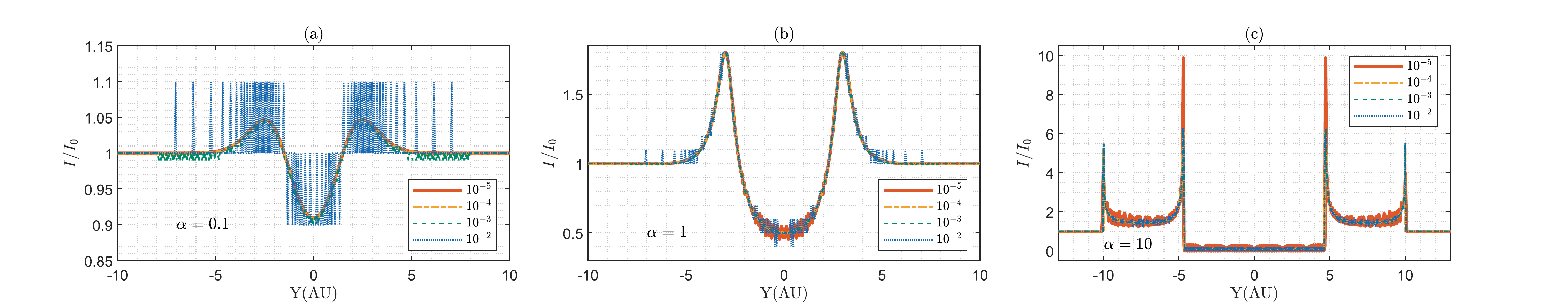}
\caption{Convergence of the plasma lensing gain with increasing resolution for different $\alpha$ values. 
The colored lines represent the gain computed at various screen resolutions, with the results converging to a stable value as the resolution improves.
The plasma lens size is $8$ au, and the screen resolution ranges from $10^{-2}$ au to $10^{-5}$ au.}
\label{Fig:ResolutionGain}
\end{figure}

Figure~\ref{Fig:ResolutionGain} illustrates the plasma lensing gain for different values of the parameter $\alpha$, with each panel corresponding to a distinct $\alpha$ value: Panel (a) for $\alpha =0.1$, panel (b) for $\alpha=1$, and panel (c) for $\alpha = 10$.
In each panel, the colored lines represent the gain computed at various screen resolutions.
Here, the plasma lens size is fixed at $8$ au, while the screen is discretized into grid elements with resolutions of $10^{-2}$ au, $10^{-3}$ au, $10^{-4}$ au, and $10^{-5}$ au. 
As the resolution increases, the results converge to a stable value. 
However, the rate of convergence and the specific behavior are contingent upon the overall system parameters, precluding a universal description of the convergence process. 
Though each case must be considered individually, it is evident that increasing the resolution ultimately leads to convergence.

\begin{figure}[ht!]
\centering
\includegraphics[width=1\textwidth]{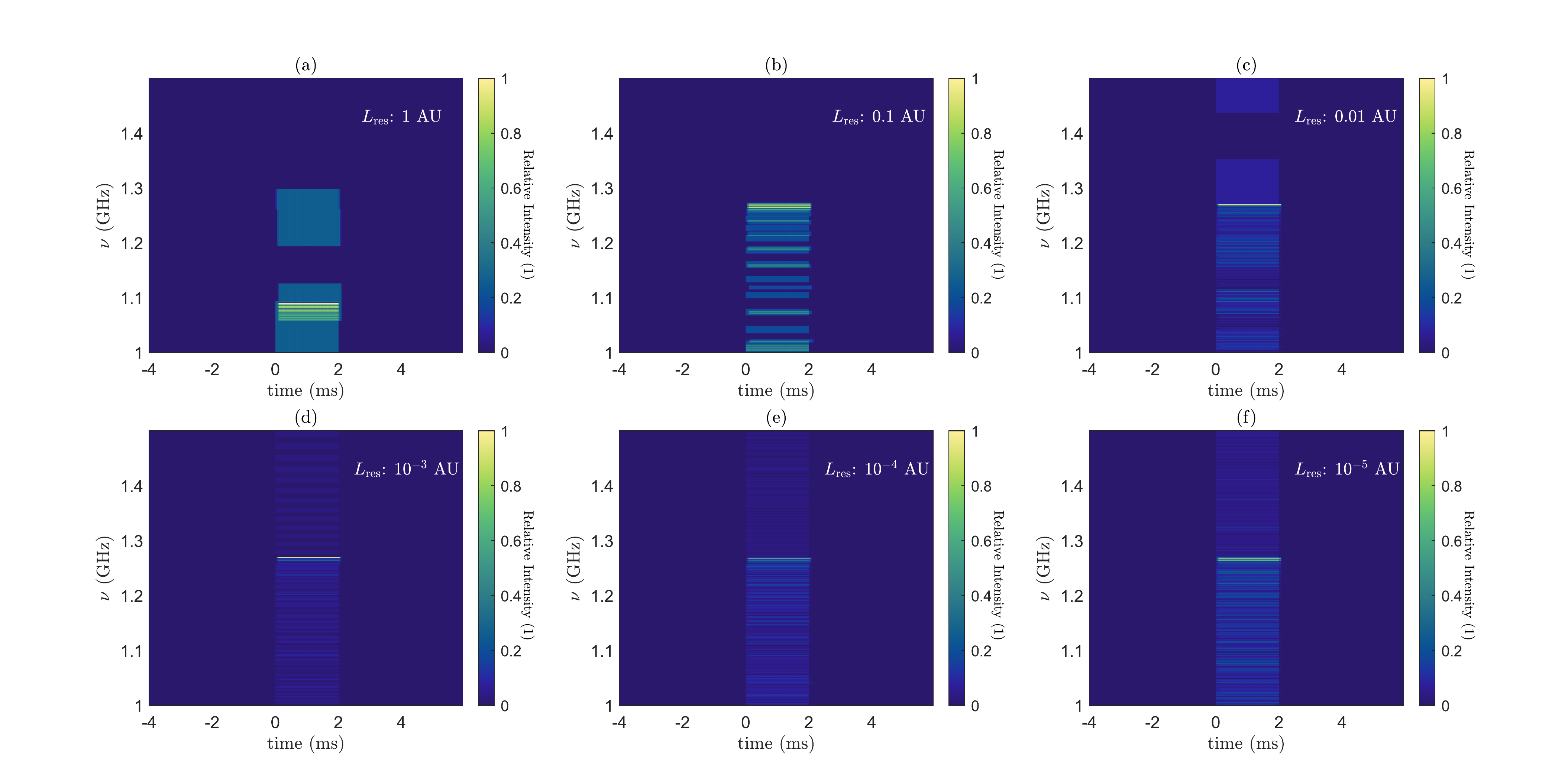}
\caption{Convergence of the received signal with increasing resolution. The figure demonstrates how the signal stabilizes as the resolution improves, with finer grids yielding sharper waterfall plots. 
The test employs a 2D Gaussian plasma lens with a peak density of $\rm 0.15 \ cm^{-3}$, a thickness of $0.1$ pc, and width of $2$ au. The $x$-aixs connects the source and the center of the screen, with both the source and the observer plane located $1000$ pc away from the screen plane. The observer is positioned $10$ au from the $x$-aixs.} 
\label{Fig:ResolutionReceive}
\end{figure}

The effect of resolution on the received signal is demonstrated in Figure~\ref{Fig:ResolutionReceive}. 
With a fixed plasma screen and observer position, the resolution is varied by adjusting the grid size from $1$ au down to $10^{-5}$ au.
At coarser resolutions, the results exhibit significant instability and large fluctuations in the signal. 
Under such low-resolution conditions, some frequency components may exhibit detectable signal arrivals while others remain undetected, leading to significant stochastic variations. 
Further more, regions accessible to photon detection tend to exhibit uniform intensity characteristics due to the limited number of ray paths contributing to the composite signal.

As the resolution increases, the simulation results progressively stabilize. Finer grids yield waterfall plots with enhanced sharpness.
Crucially, the spectral domain where caustic structures become localized near the observer displays pronounced intensity enhancement.

This stabilization confirms that resolution refinement enables precise representation of the light distribution and construction of the simulated waterfall plot, leading to a more faithful modeling of the authentic plasma lensing effect.

\subsection{Extended Results}
Expanding upon the foundational configurations established in preceding analyses, this section presents scenarios that incorporate conditions more representative of real-world astrophysical environments.
These examples provide an opportunity to examine the applicability of this code to detailed astrophysical phenomena more closely, thereby showcasing its capability to handle intricate spatial configurations and variable parameters in complex real-world scenarios.

\begin{figure}[ht!]
\centering
\includegraphics[width=1\textwidth]{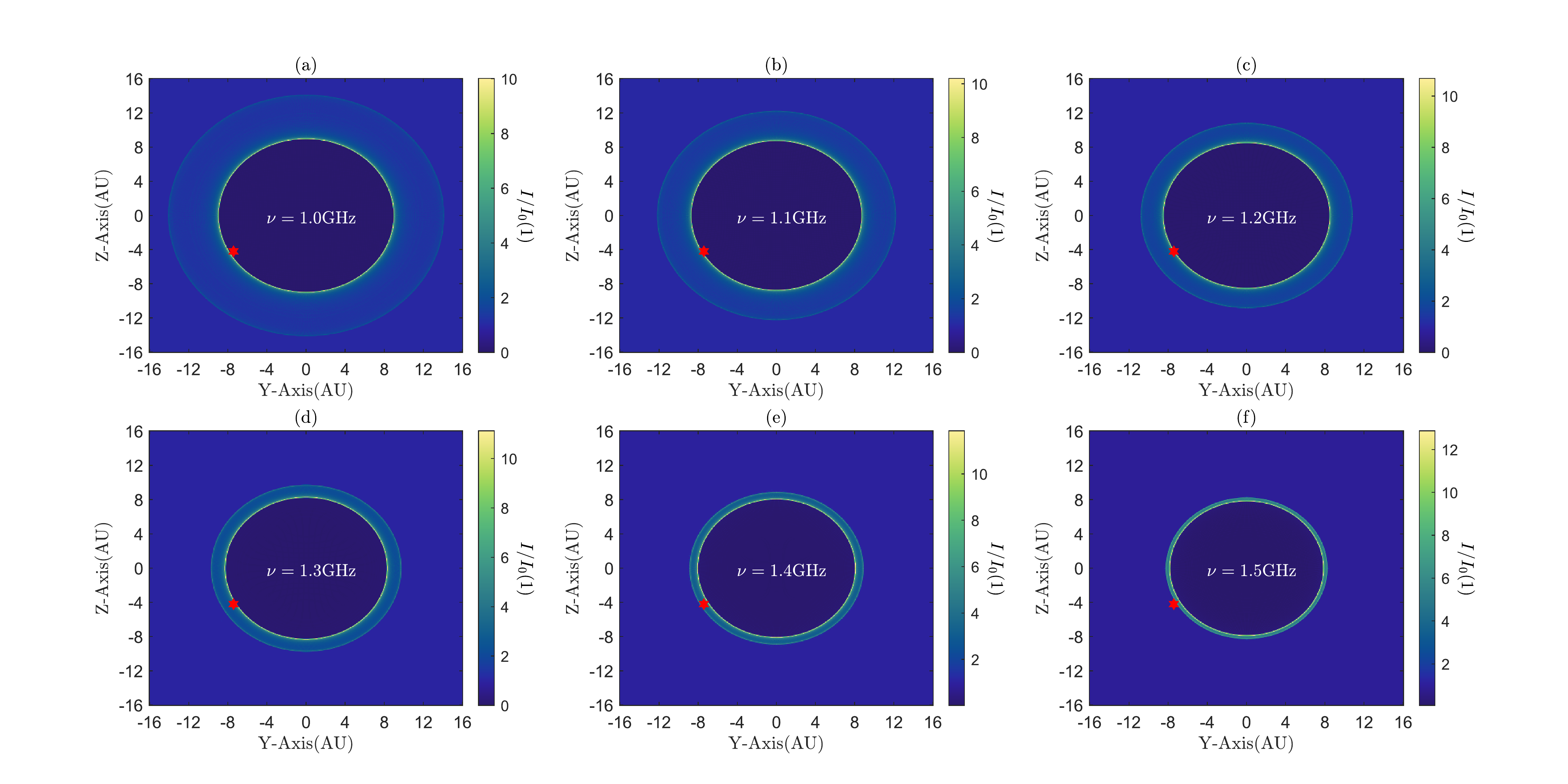}
\caption{Intensity distribution on the observer plane for a 2D Gaussian lens. The setup comprises a point source and a 2D Gaussian lens (with no variation along the x-axis) characterized by a maximum column electron density of $N_{\rm 0} = 0.015 \,\rm pc\, cm^{-3}$, and a lens size of $a = 2$ au.
Both the source-to-screen and screen-to-observer distances are set to $1$ kpc. 
The color bar indicates the relative intensity normalized by the intensity in the absence of lensing. The red star marks the location of the observer.
}
\label{Fig:Lens2D}
\end{figure}

Figure~\ref{Fig:Lens2D} displays the intensity distribution on the observer plane for a 2D Gaussian lens. The lens is modeled by the distribution
\begin{equation}
    N_{e}(y, z) = N_{0}\exp\left[-(y^{2}+z^{2})/a^{2}\right],
\end{equation}
which is analogous to the 1D Gaussian case (with $ N_{\rm e} = n_{\rm e} l $, where $ n_{\rm e} $, $l$, and $ a $ defined as in the 1D formulation).
In this example, it sets $N_{\rm 0} = 0.015 \, \rm pc\, cm^{-3}$, and $ a = 2$ au, with both the source-to-screen and screen-to-observer distances equal to $1$ kpc.
The source is modeled as a point source with a half-open angle of $10^{-5}$ degrees, ensuring that the projected circle on the screen encompasses the entire lens.
The color bar indicates the relative intensity normalized by the intensity in the absence of lensing.

\begin{figure}[ht!]
\centering
\includegraphics[width=0.7\textwidth]{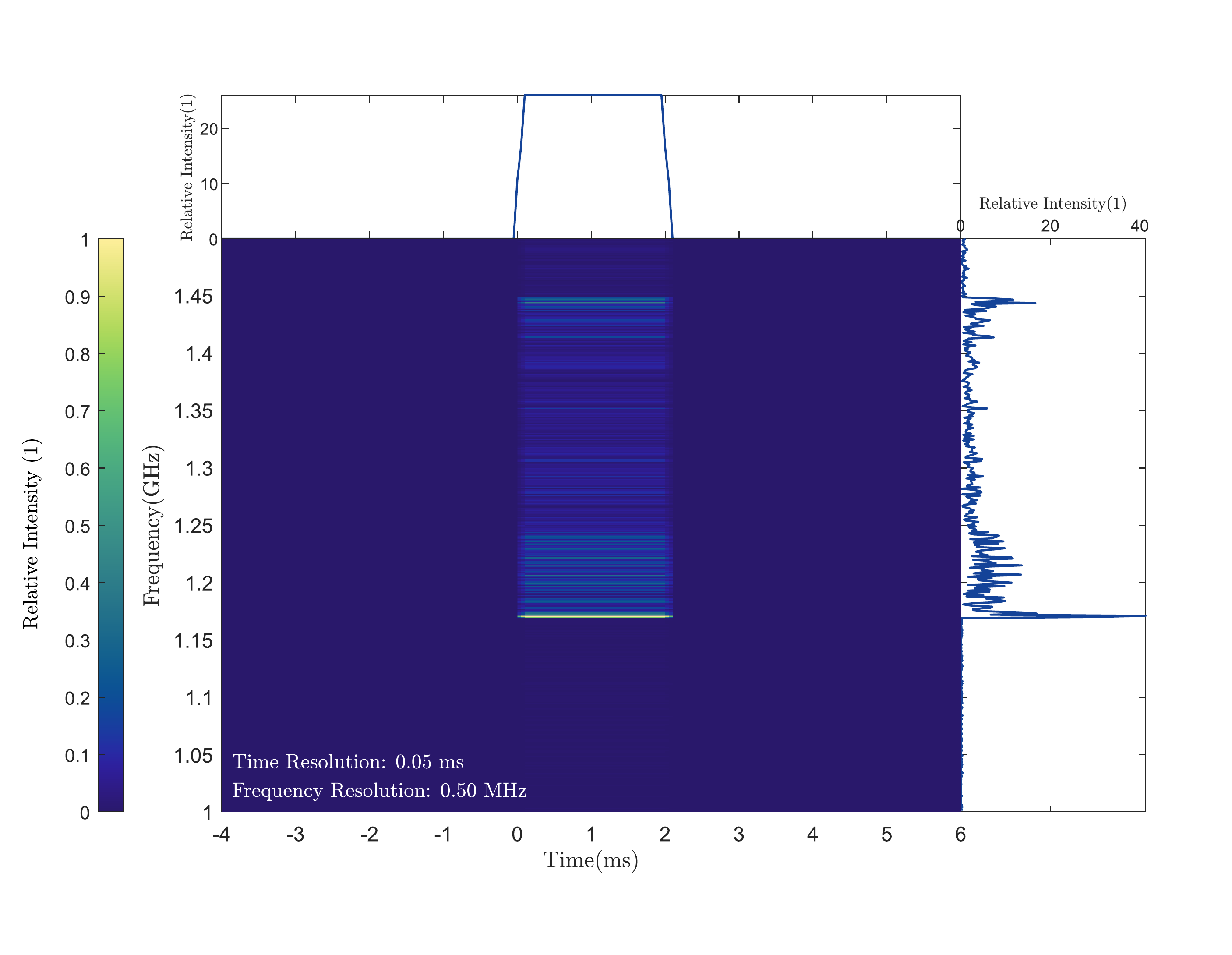}
\caption{Waterfall plot showing the observed signal at the location marked by the red star in Figure~\ref{Fig:Lens2D}.
The intensity profile reveals frequency-dependent lensing effects from a 2D Gaussian lens, with dual spectral spikes at the band edges arising from inner/outer caustic crossings, consistent with theoretical predictions by \citet{Cordes_2017}. These scintillation bands are an exclusive feature of numerical plasma lensing calculations.
}
\label{Fig:Lens2D_Receive}
\end{figure}

Generally, the lens exhibits stronger effects at lower frequencies. 
Numerical calculations demonstrate bright annular regions (caustics) with dramatically enhanced relative intensities, precisely where theoretical treatments predict divergence. 
The lens behaves as a diverging element, redistributing light from the central region toward the periphery and thereby forming two distinct caustic rings. 
In regions distant from the lens, where the electron density is nearly uniform, no lensing occurs. 
This behavior is consistent with the findings of \citet{Clegg_1998}, which indicate that inner caustics are less sensitive to variations than outer caustics.
The red star in Figure~\ref{Fig:Lens2D} marks the location of the observer, corresponding to the waterfall plots presented in Figure~\ref{Fig:Lens2D_Receive}.

Figure~\ref{Fig:Lens2D_Receive} illustrates the waterfall plots at the observer location marked by the filled red star in Figure~\ref{Fig:Lens2D}.
This plot highlights how the lensing structure alters the received signal, manifesting as variations in intensity and spectrum. 
The DM value of the plasma lens is sufficiently low and any background DM has been removed, resulting in a flat-top square wave pattern in the waterfall plot. 
Although the intrinsic duration of the signal is $2$ ms, the observer receives light from different regions of the screen with slightly varied arrival times, which results in a modest extension of the observed signal duration--an effect that becomes more pronounced at lower frequencies.
This phenomenon has the same physical origin as scattering broadening, wherein variations in path length elongate the temporal profile of the signal.

Furthermore, at lower frequencies, the observer (red star) lies within the inner caustic region (as seen in Figures~\ref{Fig:Lens2D}(a) and (b)),  where negligible light is received. 
As the frequency increases, the caustics shift toward the center and their intersection with the observer triggers abrupt intensity surges.
In the annular region between the two caustics, the intensity remains elevated relative to the nonlensed case, owing to the superposition of uniform background intensity and scattered light from the central region. Exterior to the outer caustic, the signal reverts to its normal, unaltered state. 
In the spectral domain, these effects result in the characteristic two-spike feature, as predicted by \citet{Cordes_2017}.
Additionally, interference fringes and multiscale variations emerge. 
These striped intricate patterns in the plot arise from the interference of numerous light rays with varying phases, with a weak beat-like effect--stemming from the interplay between the scintillation bandwidth and the spectral resolution--further modulating the pattern.

\begin{figure}[ht!]
\centering
\includegraphics[width=0.7\textwidth]{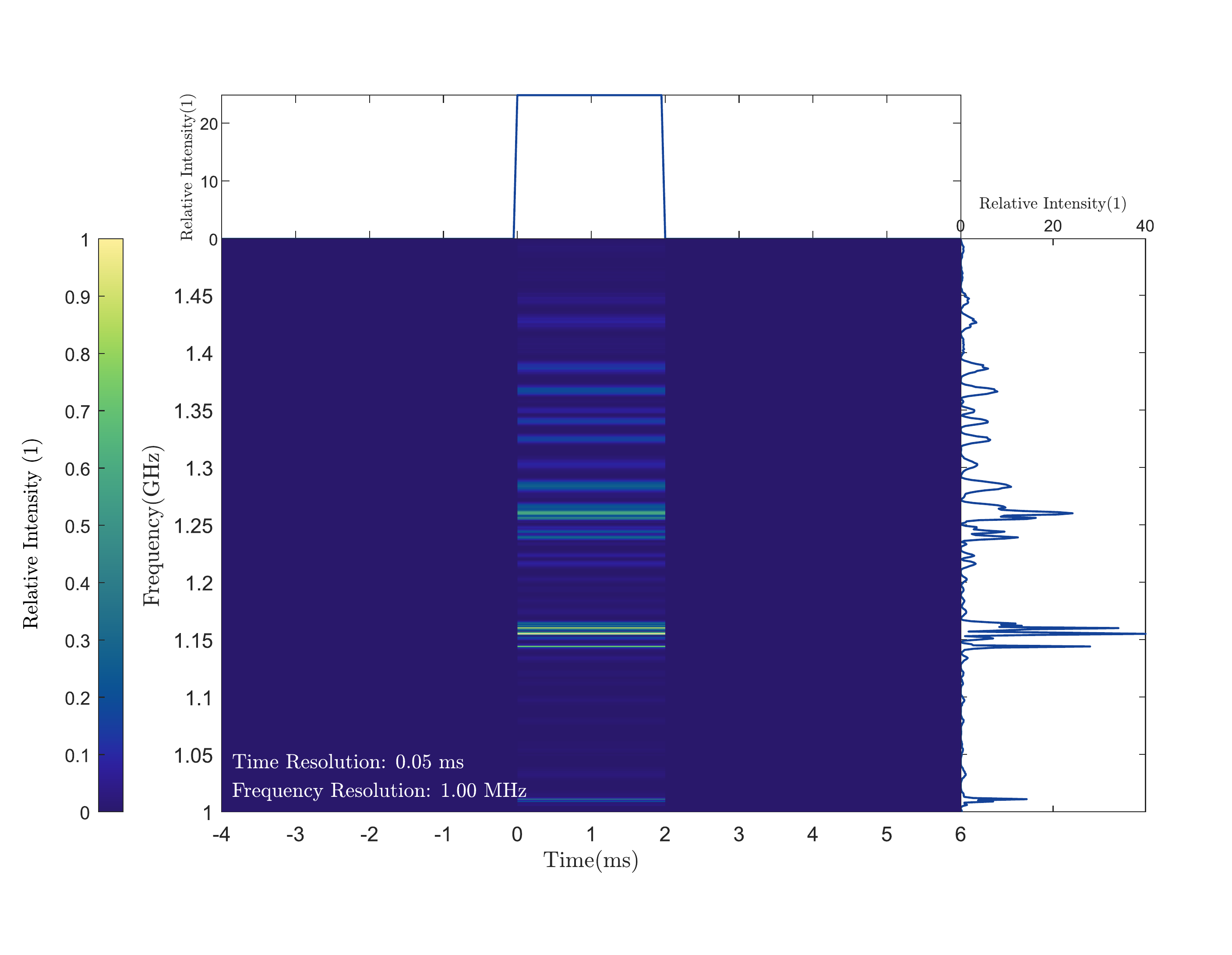}
\caption{Received signal patterns after propagation through a screen with large plasma clumps. The screen is positioned $50$ kpc from the observer and $1$ Gpc from the source, with clumps broadly distributed across its extent. 
Clumps located approximately $1$ au from the direct line connecting the source and observer effectively deflect light, producing scintillation stripes. Prominent spikes in the signal correspond to caustics generated by individual clumps.}
\label{Fig:Blob}
\end{figure}

The model generates distinct intensity spikes superimposed on the standard scintillation patterns through the redistribution of light caused by structured plasma density variations.
This phenomenon could potentially help understand the narrowband properties of FRBs. 
Macroscopic clumps, on the order of an astronomical unit, can function as plasma lenses that alter the
paths and received intensities of light.
While a single Gaussian lens typically generates a two-peak scintillation pattern in the spectrum, the presence of multiple clumps can lead to several spectral peaks at the observer. 
Unlike normal scintillation, where enhanced intensity results from constructive interference, these peaks arise from the convergence of redistributed light, thereby increasing the signal intensity at specific frequencies. 
An example of this effect is shown in Figure~\ref{Fig:Blob}.
The screen contains numerous large clumps, each approximately $1$ au in size, with a 3D Gaussian density distribution and a central maximum density of $\sim 1 \, \rm cm^{-3}$, characteristics similar to those depicted in Figure~\ref{Fig:density_example} (a). 
Each clump acts as a lens, producing a ring-like projection on the observer plane.
At a given observer location, only a subset of these clumps effectively deflects light toward the observer, thereby contributing to the final waterfall plot. In this configuration, the screen is placed $50$ kpc from the observer plane and $1$ Gpc from the source, with the contributing clumps situated roughly $1$ au from the line connecting the source and observer, resulting in scintillation stripes.
The prominent spikes in the received signal correspond to caustics produced by individual clumps.

\section{Summary and Discussion} \label{sec:Summary}
With the rapid advancements in both observations and theoretical studies in the FRB field, the demand for precise and comprehensive numerical methods to investigate propagation effects has significantly increased. 
This study presents a numerical framework, EMPI, to model the interaction between electromagnetic waves and plasma, specifically addressing linear, nonmagnetized plasma-induced propagation effects on FRBs.
The framework unifies all four types of non-magnetized plasma effects, enabling a comprehensive analysis of plasma-induced distortions in FRB signals.
The functionality and reliability of the method are rigorously validated by comparing numerical results with analytical solutions or estimates.
Additionally, we present examples involving more complex scenarios to demonstrate the capability of this framework in addressing realistic problems.
The numerical results confirm the accuracy and robustness of this code, establishing EMPI as a reliable and valuable tool for studying plasma-induced propagation effects.

As a nascent computational tool, EMPI currently exhibits three principal limitations requiring methodological refinement:
(i) As discussed in Section \ref{Sec_Regime_selection}, the regime selection mechanism lacks automated quantitative criteria, requiring researchers to align the regime with their specific objectives and evaluation criteria manually.
To address this limitation, future updates will introduce more scientifically rigorous and quantitatively defined selection criteria. 
Additionally, a module will be developed to analyze plasma density grids, enabling automatic identification of plasma structures and regime selection based on newly established, more computationally applicable criteria.
(ii) As noted in Section \ref{sec:Results}, parameter sensitivity exerts substantial influence on numerical outcomes. 
Benchmark studies demonstrate that the selection of critical parameters and grid resolution decisively governs solution accuracy.
Although standardized output protocols and operational guide lines have been established, the implementation of advanced optimization algorithms remains imperative to enhance numerical stability. 
(iii) The reconstruction of intervening plasma distribution from observational data presents fundamental challenges. 
As propagation effects arise from the combined influence of multiple physical processes, inverting observational data to determine plasma configurations constitutes a significant scientific obstacle. 
In 3D cases, this complexity escalates substantially. 
Different parameter combinations can yield nearly identical observational signatures, and a fixed plasma distribution produces markedly distinct received signals when observed from varying positions. 
Nevertheless, this approach provides a viable framework for constraining plasma parameters.
Future enhancements incorporating magnetic field effects and FRB temporal evolution characteristics will synergistically combine with additional multidimensional observational data, systematically improving parameter inversion accuracy through comprehensive constraint optimization.

Notwithstanding these limitations, the EMPI framework demonstrates transformative potential in plasma propagation research through four key advancements: 
(a) Breakthrough in 3D structural effects. 
While prior studies predominantly employed idealized 1D analytic plasma lens models, EMPI pioneers high-resolution 3D plasma distribution calculation--critical for resolving parametric degeneracies in astrophysical
plasmas. 
For instance, plasma configurations with identical column densities generate markedly different propagation signatures depending on their spatial morphology, a phenomenon quantitatively inaccessible through traditional analytic approaches. 
The primary goal of this paper is to introduce a comprehensive numerical framework that facilitates the exploration of various plasma distributions, rather than to explain the underlying physics of specific distributions. 
More complex cases and discussions will be addressed in future work. 
(b) Novel observational constraint methodology. This approach bridges the gap between idealized analytical plasma models and fully turbulent distributions characterized by their statistical proper
ties. 
By enabling more realistic numerical calculations, the EMPI framework provides a means to constrain FRB and plasma parameters by direct comparison with observations. The utilization of Markov Chain Monte Carlo methods with this framework will further refine constraints on the properties of the intervening plasma and FRBs. 
(c) Dynamic evolution modeling capability. 
The flexibility to adjust the positions of both the screen and the observer allows for the incorporation of velocity terms. 
This capability will enable calculations of scintillation arc phenomena using time-series snapshots, providing a new approach to studying FRB variability and propagation effects.
(d) Generalized astrophysical applicability. The methodology presented in this study is applicable to other scenarios where similar conditions hold.
In particular, it can be naturally extended to other radio sources, especially transients. 
If the computational method criteria are met, this approach could also be extended to other wavelengths, broadening its applicability in astrophysics.

In summary, this study establishes a robust methodology for analyzing electromagnetic wave propagation through astrophysical plasma, advancing the understanding of plasma mediated interactions. The subsequent paper in this series will focus on magnetized plasma environments, specifically analyzing how magnetic fields alter the polarization properties of electromagnetic signals.

\section{Acknowledgments}

\begin{acknowledgments}

N.X., H.G., T.-C.W., and R.G. are supported by the National SKA Program of China (2022SKA0130100), the National Natural Science Foundation of China (Projects 12373040,12021003) and
the Fundamental Research Funds for the Central Universities.

Y.-P.Y. is supported by the National Natural Science Foundation of China grant No.12473047, the National Key Research and Development Program of China (2024YFA1611603) and the National SKA Program of China (2022SKA0130100).

W.-Y. W. acknowledges support from the NSFC (No.~12261141690 and No.~12403058), the National SKA Program of China (No.~2020SKA0120100), and the Strategic Priority Research Program of the CAS (No.~XDB0550300).
\end{acknowledgments}

\appendix

\section{Derivation of the Critical Plasma Clump Scale} \label{criteria}
For a single plasma bump with size $l$,  the diffraction angle $\theta_{\rm diff}$ for a wave with wavelength $\lambda$ can be estimated using the following relation \citep{draine_physics_2011}:
\begin{equation}
\theta_{\rm diff} \approx \lambda/l.
\end{equation}
In non-magnetized plasma, the dispersion relation is given by \citep{Rybicki1991} as:
\begin{equation}
    \omega^{2} = \omega^{2}_{\rm p} + k^{2}c^{2}.
\end{equation}
Then the wave number $k$ can be expressed as:
\begin{equation}
    k = \frac{1}{c}\sqrt{\omega^{2}-\omega_{\rm p}^{2}} \approx \frac{\omega}{c} \left( 1 - \frac{\omega^{2}_{\rm p}}{2\omega^{2}} \right),
\end{equation}
where the approximation holds for radio waves, such as FRBs with frequencies around $\sim$ GHz, that satisfy the condition $\omega \gg \omega_{\rm p}$.
The phase shift experienced by the electromagnetic wave can be estimated as:
\begin{equation}
    \phi  = \int \frac{\omega}{c} \left( 1 - \frac{\omega^{2}_{\rm p}}{2\omega^{2}} \right) ds.
\end{equation}
This expression has two components: the first term corresponds to the vacuum phase, while the second term accounts for the phase change induced by plasma. So, the phase shifts as the wave passes through a plasma bump of size 
$l$ can be derived as:
\begin{equation}
\phi = l\frac{\omega_{\rm p}^{2}}{2c\omega}=\frac{q^{2}nl}{m_{\rm e}c\nu}.
\end{equation}
Thus, the refraction angle $\theta_{\rm ref}$ can be expressed as:
\begin{equation}
    \theta_{\rm ref} \approx \frac{\Delta x}{l} = \frac{\lambda \Delta \phi}{2\pi l} = \frac{q^{2}\Delta n }{2\pi m_{\rm e}\nu^{2}}.
\end{equation}
Setting $\theta_{\rm ref} = \theta_{\rm diff}$ allows us to derive the critical size $l_{\rm c}$
\begin{equation}
    l_{\rm c} = \frac{2\pi m_{\rm e}\nu c}{q^{2} \Delta n}.
\end{equation}

\bibliography{sample631}{}
\bibliographystyle{aasjournal}



\end{document}